# Self-supporting structure design in additive manufacturing through explicit topology optimization

Xu Guo[1], Jianhua Zhou, Weisheng Zhang[2], Zongliang Du, Chang Liu and Ying Liu

*State Key Laboratory of Structural Analysis for Industrial Equipment,*

*Department of Engineering Mechanics,*

*International Research Center for Computational Mechanics,*

*Dalian University of Technology, Dalian, 116023, P.R. China*

**Abstract**

One of the challenging issues in additive manufacturing (AM) oriented topology optimization is how to design structures that are self-supportive in a manufacture process without introducing additional supporting materials. In the present contribution, it is intended to resolve this problem under an explicit topology optimization framework where optimal structural topology can be found by optimizing a set of explicit geometry parameters. Two solution approaches established based on the Moving Morphable Components (MMC) and Moving Morphable Voids (MMV) frameworks, respectively, are proposed and some theoretical issues associated with AM oriented topology optimization are also analyzed. Numerical examples provided demonstrate the effectiveness of the proposed methods.

---

[1]Corresponding author.    E-mail: guoxu@dlut.edu.cn        Tel: +86-411-84707807
[2]Corresponding author.    E-mail: weishengzhang@dlut.edu.cn   Tel: +86-411-84707807

# 1. Introduction

Additive manufacturing (AM, also known as 3D printing) is a relatively new manufacture technique which enables the fabrication of components in an additive (layer-by-layer) way. In contrast to traditional subtractive and formative manufacturing techniques (e.g., machining and casting), AM has intrinsic ability to build components with very complex structural geometries and highly optimized mechanical/physical functionalities. On the other hand, topology optimization, which aims at designing innovative and lightweight products by distributing material within a prescribed domain in an optimal way, has reached a certain level of maturity and becomes a well-established research area [1-3]. Up to now, many methods have been proposed for topology optimization and some of them have already been applied successfully in various application fields.

Although topology optimization has great potential to become a perfect design tool that can fully exploit the tremendous design freedom provided by AM, it must be admitted that existing topology optimization approaches cannot be fully adapted to the current AM techniques. This is because although seemingly a free-form manufacturing technique, AM does have some design limitations which must be taken into account when AM oriented topology optimization approaches are devised for making topology optimization and AM an ideal fit. These limitations can be summarized briefly as follows (see also Fig. 1 for a schematic illustration). Firstly, since 3D printing is achieved either by depositing material (in a Fused Deposition Modeling, FDM), applying focused laser to cure powder (in Selective Laser Melting, SLM), spraying liquid binding onto particles (in Inkjet Printing, IP) or by using some combinations of these approaches, the achievable smallest print resolution will inevitably be influenced by the parameters such as the nozzle diameter (in FDM) or the beam width/offset in laser sintering [4]. Therefore, optimal designs with too small feature sizes (e.g., the wall thickness or the diameter of a void) may not be fabricated even with AM technique. Secondly, since in AM products are actually being built in a layer by layer way, each part of the products must be sufficiently supported from below during the manufacturing process otherwise the quality of the product cannot be fully guaranteed. Usually, support material must be introduced to manufacture certain structural topologies in order to prevent the structural material from being distorted too much or even fail (due to high bending stresses) during the build process. This treatment, however, will waste print time, increase material usage and/or require chemical processing to remove the support material. It is therefore highly desirable to design structures that are self-supporting during the course of AM. Thirdly, the inclination of downward facing and non-supported (overhang) surfaces cannot be at too large an angle with respect to the build direction. The component with a high overhang angle may deform, droop or warp during the printing process. The upper bound of the overhang angle is obvious material and process-dependent and has been investigated extensively in literature. A commonly accepted value of the maximum overhang angle amounts to $40°$-$50°$ [5, 6]. Lastly, there should be no enclosed voids existing in the structure otherwise it will be very difficult to get the unmelted powder (in SLM) and support material (in FDM) out of the void once the structure is built up by AM. Based on the above discussions, it can be concluded that if topology optimization is expected to be used as a tool for innovative design purpose, the above issues must be taken into consideration in order to guarantee that the corresponding products/structures can be successfully manufactured with AM. The readers are also referred to Brackeet et al. [7, 8] for an overview of the corresponding issues in AM oriented topology optimization.

Recent years witnessed an increasing interest to develop AM oriented topology optimization approaches by taking the aforementioned manufacturing constraints into consideration. In order to control the minimum length scale in topology optimization, both global implicit [9-11] and local explicit [12-17] methods are developed. We refer the readers to the above works and the references therein for more recent advances on this aspect. By introducing a virtual temperature field, Li et al. developed an approach which can deal with connectivity constraint in topology optimization and produce optimal designs without enclosed voids [18]. In order to tackle the issues of designing of self-supporting structures and controlling overhang features, Leary et al. [19] and Hu et al. [20] proposed the post-processing methods to ensure the printability without introducing additive support materials. This is achieved by modifying the theoretically optimal topologies to respect the overhang angle constraint. In [7], the authors suggested first identifying the angles of overhang parts in structures obtained by topology optimization and then achieving the overhang angle control by incorporating the corresponding constraints in optimization problem explicitly. No subsequent research work following this idea, however, can be found in literature. Langelaar [21, 22] proposed an AM oriented topology optimization formulation to design self-supporting optimal structures without additional supporting materials. To this end, the author developed a filter scheme that can incorporate the main characteristics of a typical AM process and implemented it in a density based topology optimization approach. It was reported that fully self-supporting can be obtained with use of this approach. More recently, Gaynor and Guest [23] developed a topology optimization approach to produce self-supporting optimal designs. In this approach, a series of projection operators that can enforce the minimum length scale constrains and restrict the overhang angles are introduced under the variable density solution framework. Numerical examples showed that self-supporting optimal designs satisfying minimum length scale, overhang angle and volume constraints do can be obtained with use of this approach.

Although numerous efforts have been made to resolve the aforementioned printable issues associated with AM processes, there is still room for further improvement. For example, the post-processing methods suggested in [19] and [20] will inevitably introduce extra computational efforts and more importantly destroy the optimality of the original optimized designs. Although the strategy of introducing additional structures to a previously optimized design [19, 20] does can make it self-supporting, the mass of the structure, however, will increase substantially and the structure after this treatment may deviate significantly from the originally optimized geometry [21]. The problem associated with the filter approach invented in [21] and [22] is that the filter functions introduced are highly nonlinear and dependent on the regular mesh used for finite element analysis. It is also found that there exist a relatively large number of elements with intermediate densities in some optimized designs provided. This phenomenon is obvious the side effect of the corresponding filter operation. It is also unclear how to optimize the build orientation in the proposed solution approach. For the projection approach developed in [23], as pointed out by the authors, since the determination of structural topology and the corresponding sensitivity analysis must be carried out in a layer-by-layer manner, the proposed approach is in general computationally inefficient. Furthermore, for some problems, the involvement of multiple embedded nonlinear functions in the solution scheme may also lead to convergence issues.

Compared to adding extra material or carrying out post-processing to make an optimal structure printable, it is generally believed that designing self-supporting printable structures through topology optimization method directly is more preferable since it can simplify the post-processing and reduce the manufacture cost. In the present work, we intend to discuss how to design self-supporting structures produced by AM without introducing support materials in a more explicit and geometrical way. Our motivation is from the consideration that since the constraints associated with the designing of self-supporting structures (e.g., overhang angle, minimum length scale) are actually geometrical in nature, it seems more appropriate to include more geometry information in the mathematical formulation of the considered problem and perform topology optimization in an geometrically explicit way (it is worth noting that topological design is usually achieved in an implicit way in tradition solution approaches such as variable density method and level set method, see [24-29] for more detailed discussions on this aspect). Recent years witnessed a growing interest in solving topology optimization problems by optimizing a set of geometrical parameters explicitly (i.e., a revival of shape optimization) [17, 24-29]. In particular, the so-called Moving Morphable Components (MMC) and Moving Morphable Voids (MMV) where a set of components (in MMC) or voids (in MMV) are used as basic building blocks of optimization have been developed. As demonstrated in the following sections, the MMC and MMV approaches can deal with the problem of designing self-supporting structures in a more natural way since more geometry information (e.g., the outward normal vector of structural boundary, the inclined angle of a structural component) is embedded in the corresponding problem formulations. In the present work, two approaches for designing AM oriented self-supporting structures established in the MMC and MMV solution framework are developed, respectively.

The rest of the paper is organized as follows. In Section 2, the proposed two solution approaches are presented and analyzed in detail. Some theoretical issues associated with AM oriented topology optimization are discussed in Section 3. Numerical solution aspects are addressed in Section 4. In Section 5, some numerical examples are presented to demonstrate the effectiveness of the proposed methods. Finally some concluding remarks are provided in Section 6.

## 2. Two approaches for designing self-supporting structures

In this section, two approaches for designing self-supporting structures established based on the Moving Morphable Components (MMC) and Moving Morphable Voids (MMV), respectively, will be presented. As a preliminary attempt to address this difficulty problem, only two dimensional problems are considered in the present work.

**2.1 MMC-based approach**

As discussed in the previous section, a critical issue in topology optimization of self-supporting structures is to control the inclined angles of structural components in the structure. Under this circumstance, the MMC-based explicit topology optimization approach first initialized in [24] where the inclined angles of each component are adopted as design variables, is a natural choice to serve this purpose. In the following, a brief introduction of the basic idea of MMC will be given first. As pointed out in [24], unlike in traditional topology optimization approaches where structural topologies are represented either by element densities (in variable density approach) or by nodal values of a level set function (in level set approach), in the MMC-based approach, a set of moving morphable components are adopted as basic building blocks of topology optimization (see Fig. 2 for a schematic illustration). These components are allowed to move, deform, overlap and merge in the design domain freely, and optimal structural topology can be obtained by optimizing the positions, inclined angles, lengths, widths and the layout of these components. This treatment provides a new paradigm for topology optimization and has big potential to resolve some challenging issues that cannot be dealt with easily by traditional methods. We refer the readers to [24, 25, 27-29] for more details on the generalization and variants of this approach.

In the present work, as a preliminary attempt to generate self-supporting structures through topology optimization, we propose to use structural components with hyperelliptic shapes as building blocks of topology optimization [24]. Under this circumstance, the corresponding topology optimization problem can be formulated as:

$$\text{Find} \quad \boldsymbol{D} = ((\boldsymbol{D}^1)^\top, \ldots, (\boldsymbol{D}^{nc})^\top)^\top, \; \alpha \; \text{and} \; \boldsymbol{u}$$
$$\text{Minimize} \quad I = I(\boldsymbol{D}, \boldsymbol{u})$$
$$\text{S.t.}$$

$$\int_D H(\chi^s)\mathbb{E}^s : \varepsilon(\boldsymbol{u}) : \varepsilon(\boldsymbol{v}) \mathrm{d}V = \int_D H(\chi^s)\boldsymbol{f} \cdot \boldsymbol{v} \mathrm{d}V$$

$$+ \int_{\Gamma_t} \boldsymbol{t} \cdot \boldsymbol{v} \mathrm{d}S, \quad \forall \boldsymbol{v} \in \mathcal{U}_{ad},$$

$$V = \int_D H(\chi^s) \mathrm{d}V \leq \bar{V} \mathrm{meas}(D),$$

$$(\sin(\theta_k + \alpha))^2 \geq (\sin(\bar{\theta}))^2, \quad k = 1, \ldots, nc,$$

$$\boldsymbol{D} \subset \mathcal{U}_{\boldsymbol{D}},$$

$$0 \leq \underline{\alpha} \leq \alpha \leq \bar{\alpha} \leq \pi/2,$$

$$\boldsymbol{u} = \bar{\boldsymbol{u}}, \quad \text{on } \Gamma_u. \tag{2.1}$$

In Eq. (2.1)

$$\chi^s(\boldsymbol{x}) = \max(\chi^1(\boldsymbol{x}), \chi^2(\boldsymbol{x}), \cdots, \chi^{nc}(\boldsymbol{x})) \tag{2.2}$$

is the topology description function (TDF) of the region occupied by the components and $\chi^k$ ($k = 1, \cdots, nc$) denotes the TDF of the region occupied by the $k$-th component (i.e., $\Omega_k$), that is,

$$\begin{cases} \chi^k(\boldsymbol{x}) > 0, & \text{if } \boldsymbol{x} \in \Omega_k, \\ \chi^k(\boldsymbol{x}) = 0, & \text{if } \boldsymbol{x} \in \partial\Omega_k, \\ \chi^k(\boldsymbol{x}) < 0, & \text{if } \boldsymbol{x} \in D\setminus\Omega_k. \end{cases} \tag{2.3}$$

and

$$\chi^k(x,y) = 1 - \left(\frac{x'}{L_k}\right)^p - \left(\frac{y'}{t_k}\right)^p, \tag{2.4}$$

with

$$\begin{Bmatrix} x' \\ y' \end{Bmatrix} = \begin{bmatrix} \cos\theta_k & \sin\theta_k \\ -\sin\theta_k & \cos\theta_k \end{bmatrix} \begin{Bmatrix} x - x_{0_k} \\ y - y_{0_k} \end{Bmatrix}, \tag{2.5}$$

where $p$ is a relatively large even number (we take $p = 6$ in the present study). Actually Eq. (2.4) represents a hyperelliptic shape component centered at point $(x_{0_k}, y_{0_k})$ with a half-length $L_k$, a half-thickness $t_k$ and a inclined angle $\theta_k$ (with respect to the horizontal axis). The vector of design variables associated with the $k$-th component is $\boldsymbol{D}^k = (x_{0_k}, y_{0_k}, L_k, t_k, \theta_k)^\top$ (see Fig. 3 for reference). Obviously,

$$\begin{cases} \chi^s(\boldsymbol{x}) > 0, & \text{if } \boldsymbol{x} \in \Omega^s, \\ \chi^s(\boldsymbol{x}) = 0, & \text{if } \boldsymbol{x} \in \partial\Omega^s, \\ \chi^s(\boldsymbol{x}) < 0, & \text{if } \boldsymbol{x} \in D\setminus\Omega^s, \end{cases} \tag{2.6}$$

where $\Omega^s = \bigcup_{k=1}^{nc} \Omega_k$ represents the region occupied by the structural components. In Eq. (2.1), the symbol $nc$ denotes the total number of components in the design domain. $\mathcal{U}_D$ is the admissible set that $\boldsymbol{D}$ belongs to. The symbols $\boldsymbol{u}$ and $\boldsymbol{v}$ are the displacement field and the corresponding test function defined on $\Omega = \bigcup_{k=1}^{nc} \Omega^k$ with $\mathcal{U}_{\text{ad}} = \{\boldsymbol{v} | \boldsymbol{v} \in \mathbf{H}^1(\Omega), \boldsymbol{v} = \boldsymbol{0} \text{ on } \Gamma_u\}$. The symbol meas(D) stands for the measure of the design domain D. The symbols $\boldsymbol{f}$ and $\boldsymbol{t}$ denote the body force density and the surface traction on Neumann boundary $\Gamma_t$, respectively. $\bar{\boldsymbol{u}}$ is the prescribed displacement on Dirichlet boundary $\Gamma_u$. The symbol $\boldsymbol{\varepsilon}$ denotes the second order linear strain tensor. $\mathbb{E}^S = E^s/(1+v^s)[\mathbb{I} + v^s/(1-2v^s)\boldsymbol{\delta}\otimes\boldsymbol{\delta}]$ ($\mathbb{I}$ and $\boldsymbol{\delta}$ represent the symmetrized fourth and the second order identity tensor, respectively) is the fourth order isotropic elasticity tensor of the solid material with $E^s$ and $v^s$ denoting the corresponding Young's modulus and Poisson's ratio, respectively. The symbol $\bar{V}$ denotes the upper bound of the relative available volume of solid material and $H = H(x)$ in Eq. (2.1) is the Heaviside function.

It is worth noting that compared to MMC-based topology optimization formulations where no self-supporting requirement is considered, the only extra constraints in the present formulation is $(\sin(\theta_k + \alpha))^2 \geq (\sin(\bar{\theta}))^2$, $k = 1, ..., nc$ where $\alpha$ is the rotation angle of the work plane (see Fig. 4 for a schematic illustration) and $\bar{\theta}$ is the lower bound of the overhang angle (as discussed before usually $\bar{\theta} \in [40°\pi/180°, 50°\pi/180°]$), respectively. These constraints are included to ensure that the inclined angles of components will not be less than the critical value (i.e., $\bar{\theta}$) which can make the components self-supportable. It is also worth noting that including the angle of working plane as a design variable increases the design freedom substantially. As shown in Fig. 4 and the examples in the following section, some optimal structures which are not printable for a specific working plane can be produced without any difficulty if the angle of working plane can be selected appropriately.

As can be seen from Eq. (2.1), the advantage of the present formulation is that the self-supporting requirement can be dealt with by introducing a set of geometry constraints explicitly. Furthermore, both the structural topology and the angle of working plane can be optimized in a simultaneous way. It is, however, should be pointed out that this formulation has a deficiency such that it cannot totally exclude the unprintable case of V-shape material distribution as shown in Fig. 5. As plotted in Fig. 5, the two components actually have no supports from below and therefore cannot be printed even though their inclined angles are far beyond the threshold value. An accompanying case is also shown in Fig. 5 where a set of overlapping components with high inclined angles may constitute an unprintable shallow overhang part of a structure. Although these cases are hypercritical and may be avoided to a large extent by optimizing the rotation angle of the working plane, a theoretically complete way to eliminate these unpleasant cases is to introduce the following pointwise supportable constraint and inclined angle constraint into the problem formulation:

$$\text{meas}\left(\Omega^s \cap C_\epsilon(\boldsymbol{x})\right) > \delta, \quad \forall \boldsymbol{x} \in \partial\Omega^s, \tag{2.7a}$$

and

$$\boldsymbol{n} \cdot \boldsymbol{b}_p \leq \cos\bar{\theta}, \quad \forall \boldsymbol{x} \in \partial\Omega^s, \tag{2.7b}$$

where $C_\epsilon(\boldsymbol{x})$ represents a semicircle of radius $\epsilon$ center on $\boldsymbol{x}$ and $\epsilon$ as well as $\delta$ are two small positive values. The symbols $\boldsymbol{n}$ is the *inward* normal vector of $\partial\Omega^s$ and $\boldsymbol{b}_p$ is the unit vector representing the print direction. We refer the authors to Fig. 6 for a schematic illustration of the implication of these geometrical constraints.

Although the constraint in Eq. (2.7) is indeed numerically implementable (actually $\text{meas}\left(\Omega^s \cap C_\epsilon(\boldsymbol{x})\right) = \int_D H\left(\min\left(\chi^s(\boldsymbol{x}), \chi_{C_\epsilon}(\boldsymbol{x})\right)\right) dV$, where $\chi_{C_\epsilon}$ is the TDF of the region occupied by $C_\epsilon(\boldsymbol{x})$), we do not intend to discuss this treatment in detail in the present study. In order to circumvent the aforementioned difficulties, another approach to generate self-supporting printable structures based on the MMV framework will be described in the next subsection.

**2.2 MMV-based approach**

In this section, we shall discuss how to design self-supporting structures suitable for AM with use of topology optimization under the so-called Moving Morphable Voids (MMV) framework through explicit boundary evolution. The central idea to introduce printable features (voids), whose boundaries can be described explicitly by a set of B-spline curves, into problem formulation and transfer the corresponding topology optimization into a shape optimization problem. As shown in Fig. 7, in the MMV-based approach, the topological change of a two dimensional structure is achieved by the deformation, intersection and merging of a set of closed parametric curves (i.e., $C_i$, $i = 1, ..., n_v$) which represent the interior boundary of the structure (i.e., boundaries of a set of voids). Unlike the traditional level set approach (which is also a boundary-based approach for topology optimization), in the MMV-based approach, there is no need to introduce an extra level set function defined in a higher dimensional space to represent the structural boundary implicitly. The design variables involved in the MMV-based approach are only the coordinates of the control points (or the related parameters) of the parametric curves. The readers are referred to [29] for more discussions on technical details of this approach.

In order to circumvent the problems associated with the MMC-based approach discussed in the previous subsection, we suggest introducing a set of printable voids with explicit boundary representation as the basic building blocks of topology optimization. In the present approach, B-spline curve expressed in Eq. (2.8) is used to describe the shape of each structural component (see Fig. 8 for reference):

$$\boldsymbol{C}(u) = \sum_{k=0}^{n} N_{k,p}(u) \boldsymbol{P}_k, \quad a \leq u \leq b, \tag{2.8a}$$

with

$$\boldsymbol{P}_i = \left(x_c + \sin\left((i-1)\theta + \frac{\pi}{2}\right)d_i, y_c + \cos\left((i-1)\theta + \frac{\pi}{2}\right)d_i\right)^\top, \quad \theta = \frac{\pi}{n-2},$$
$$i = 1, \ldots, n-1, \tag{2.8b}$$

and

$$\boldsymbol{P}_0 = \boldsymbol{P}_n = (x_c, y_c + d_0)^\top. \tag{2.8c}$$

The meanings of $x_c, y_c$ and $d_i \geq 0, i = 0, \ldots, n$ are self-evident from Fig. 8. As pointed out in [29], under this treatment, it can always guarantee that there is no cusp on the B-spline curve and the curve cannot be self-intersected. It can be seen clearly from Fig. 8 that the printability requirement can definitely be respected once the condition $d_0/d_1 \geq \tan\bar{\theta}$ and $d_0/d_3 \geq \tan\bar{\theta}$ are satisfied. Fig. 9 also plots some other printable features whose shape can be represented by the B-spline curved expressed in Eq. (2.8). It can be seen from this figure that printable features with fairly complex shapes do can be represented by the proposed geometry representation scheme. Furthermore, it is also worth noting that the V-shape issue mentioned previously, which *cannot* be account for by only imposing the inclined angle constraint (i.e., $\boldsymbol{n} \cdot \boldsymbol{b}_p \leq \cos\bar{\theta}$) can be dealt with in an easy way by introducing the printable features.

In the present study, in order to preserve the printability of each the void in the structure, it is required that every two voids cannot be intersected (i.e., $\Omega^{V_i} \cap \Omega^{V_j} = \emptyset$ where $\Omega^{V_i}$ and $\Omega^{V_j}$ are the regions enclosed by $C_i$ and $C_j$, respectively) otherwise the printability requirement may not be fully respected as shown in Fig. 10. Furthermore, we also need a B-spline curve to describe the exterior boundary of the structure (i.e., $C_0 \cap D$ in Fig. 10). In the present study, it is assumed that the central point of $C_0$ (i.e., $x_c^0, y_c^0$) is fixed, only $d_1^0, \ldots, d_{n_0}^0$ are adopted as design variables.

Obviously, the printability condition associated with the exterior boundary is $\boldsymbol{n}^{C_0} \cdot \boldsymbol{b}_p \leq \cos\bar{\theta}, \forall \boldsymbol{x} \in C_0 \cap D$, where $\boldsymbol{n}^{C_0}$ is the *inward* normal vector of $C_0$ and $\boldsymbol{b}_p$ is the unit vector representing the print direction. It is also worth noting that even though the every pair of interior boundary curves $C_i$ and $C_j$ ($i \neq j$) cannot be intersected with each other, they, however, all can be intersected with the exterior boundary curve $C_0$ as shown in Fig. 10.

At this position, it is worth noting that if only one control point at the lower part of a B-spline curve is used to construct an interior boundary, it can always guarantee that the corresponding void is printable. If, however, more control points are introduced to describe the shape of the interior boundary, as shown in Fig. 11a, it is possible that the corresponding enclosed void is not printable. This problem can be resolved by requiring that $x_i \geq x_{i+1}$, and $y_i \leq \frac{y_1 - y_0}{x_1 - x_0}(x_i - x_0) + y_0$ as well as $y_i \leq \frac{y_{n-1} - y_0}{x_{n-1} - x_0}(x_i - x_0) + y_0$, for all $i = 2, \ldots, n-2$, respectively (see Fig. 11b for reference). In the above statement $x_i$ and $y_i, i = 2, \ldots, n-1$ are the horizontal and vertical coordinates of the control points of the corresponding B-spline curve. These constraints are linear in nature and thus easy to be dealt with in numerical implementation.

Based on the above discussions, the problem formulation for design self-supporting structures under MMV framework can be written as

Find $\quad \boldsymbol{D} = ((\boldsymbol{D}^0)^\top, (\boldsymbol{D}^1)^\top, \ldots, (\boldsymbol{D}^{nv})^\top)^\top$

Minimize $\ I = I(\boldsymbol{D})$

S.t.

$\quad \mathrm{meas}\left((D \cap \Omega^{V_0}) \backslash \cup_{i=1}^{nv} \Omega^{V_i}\right) \leq \bar{V} \mathrm{meas}(D),$

$\quad \Omega^{V_i} \cap \Omega^{V_j} = \emptyset, \quad i \neq j, i, j = 1, \cdots, nv,$

$$d_0^i/d_1^i \geq \tan\bar{\theta}, \qquad i = 1, \ldots nv,$$

$$d_0^i/d_{n-1}^i \geq \tan\bar{\theta}, \qquad i = 1, \ldots nv,$$

$$\boldsymbol{n}^{C_0} \cdot \boldsymbol{b}_p \leq \cos\bar{\theta}, \qquad \forall\, x \in C_0 \cap \mathrm{D},$$

$$\boldsymbol{D} \subset \mathcal{U}_{\boldsymbol{D}}, \tag{2.9}$$

where $\boldsymbol{D}^i = \left(x_c^i, y_c^i, d_0^i, \ldots, d_{n_i-1}^i\right)^\top, i = 0, \ldots, nv$ denote the vector of design variables associated with $C_i$, resepectively. The symbol $\mathrm{meas}\left((\mathrm{D} \cap \Omega^{V_0})\backslash \cup_{i=1}^{nv} \Omega^{V_i}\right)$ represents the measure of the region $(\mathrm{D} \cap \Omega^{V_0})\backslash \cup_{i=1}^{nv} \Omega^{V_i}$ (i.e., the solid part of the structure), where $\Omega^{V_0}$ represents the solid region occupied by $C_0$.

## 3. The optimality of a self-supporting structure-some theoretical considerations

Although optimal design of self-supporting structures has been investigated intensively in literature, it is still an open question what the optimal structure should like when self-supporting requirement is taken into consideration. In this section, the optimality of a self-supporting structure will be discussed from a theoretical point of view. Here the objective and constraint functionals are taken as the compliance and the volume of solid material, respectively.

Assuming that an optimal structure obtained with topology optimization without considering self-supporting requirement takes the form shown in Fig. 12a. It can be conjectured that the corresponding optimal self-supporting solution can be constructed from it as the way shown in Fig. 12b. The key point of this treatment is that we can introduce supporting materials with infinitesimal small amount of volume and stiffness to satisfy the self-supporting requirement without degrading the stiffness of the structure. This can be explained as follows.

Without loss of generality, assuming that all void parts of an optimal structure are $l \times 1$ rectangles as shown in Fig. 12c. If we reinforce each void using supporting material as the way shown in Fig. 12d, it can be confirmed that the resulting structure is obvious self-supporting. Under this circumstance, it can be estimated that the total volume of the support material in each void is

$$V^s = \frac{l}{\delta}\left(\frac{1}{2}\delta^2 + \delta^2(1-\delta)\right) = \frac{3}{2}\delta l - \delta^2 l \propto O(\delta). \tag{3.1a}$$

Furthermore, the Voigt upper bound of the modulus of the equivalent elasticity tensor of the

*Computer Methods in Applied Mechanics and Engineering, under review*  2016-11-12  12

reinforced void can be estimated as

$$\|\bar{\mathbb{C}}^s\| \cong \frac{V^s}{V} E^s = \frac{\left(\frac{3}{2}\delta l - \delta^2 l\right)}{l \times 1} E^s = \left(\frac{3}{2}\delta - \delta^2\right) E^s \propto O(\delta E^s). \tag{3.1b}$$

From Eq. (3.1), it yields that $V^s \to 0$ and $\|\bar{\mathbb{C}}^s\| \to 0$ as $\delta \to 0$. Therefore it can be concluded that $V^\delta \to V^0, \boldsymbol{u}^\delta \xrightarrow{w} \boldsymbol{u}^0$ (weak convergence) as $\delta \to 0$ (in an appropriate function space, e.g., $H^1(D)$, where $V^\delta(V^0)$ and $\boldsymbol{u}^\delta(\boldsymbol{u}^0)$ denote the total volume and displacement field of the reinforced structure (original structure), respectively. Since the structural compliance is weakly continuous with respect to the displacement field, then it yields that $I^\delta \to I^0$ as $\delta \to 0$, where $I^\delta$ and $I^0$ denote the compliance of the reinforced structure and original structure, respectively. Although the above conclusions are obtained under the assumption that the voids in the structure are of rectangle shapes, they still hold when the structure has curved interior and exterior boundaries although the corresponding rigorous mathematical proof needs more technical treatments. It is also worth noting that the above treatment for obtaining self-supporting structures is not unique, other treatments (for example distributing the support material in a hierarchical way as shown in Fig. 12e) are also applicable and the essential feature of the corresponding mathematical analysis is the same as that of the analysis presented above.

From the above analysis, it seems appropriate to conjecture that the theoretically optimal self-supporting structure can be constructed from the corresponding optimal structure without considering the self-supporting requirement by introducing *infinitely many* rods with *infinitely small* cross sections as additional supporting structures. If no regularization technique is introduced (e.g., imposing minimum length scale constraint or total perimeter constraint, etc.), it can be expected that the numerical solution results may be highly mesh-dependent if the solution algorithms are smart and robust enough to find global optimal solutions under prescribed finite element mesh. This may also explain why regions with intermediate densities are prone to exist when variable density approach are employed to design self-supporting structures [21, 23]. Of course, if regularization formulations/techniques are introduced in prior in problem formulation or employed in the numerical solution process, these "chattering designs" can definitely be suppressed. Although the above theoretical consideration is not mathematically rigorous, it provides useful insight into the problem under consideration and may give an estimation on the lower bound of the optimal value of the objective functional.

## 4. Numerical solution aspects

In this section, we shall discuss several relevant issues for the numerical implementation of the proposed approaches.

### 4.1 Finite element analysis

In the present study, the X-FEM method proposed in [30] is employed to carry out structural response analysis based on fixed finite element (FE) mesh. Under this circumstance, it is necessary to construct a topological description function (TDF) $\chi^s = \chi^s(x)$ characterizing solid part of the structure implicitly. For the MMC-based approach, $\chi^s$ can be constructed easily since the TDF of each component (i.e., $\chi^k, k = 1, \ldots, nc$) has closed form analytical expression and $\chi^s(x) = \max(\chi^1(x), \chi^2(x), \cdots, \chi^{nc}(x))$. For the MMV-based approach, the TDF of $\chi^s$ can be constructed in the following way (see Fig. 13 for reference). Firstly, the MATLAB function **Inpolygon** and the explicit expressions of the B-spline boundary curves can be used to determine whether a FE node (e.g., $N_i$) is inside or outside the region $\Omega_j$ enclosed by a specific B-spline boundary curve $C_j$. If $N_i$ is inside $\Omega_j$ then let $\text{IN}(N_i) = 1$ otherwise let $\text{IN}(N_i) = 0$. Secondly, calculating the minimum distance $d^i$ of $N_i$ to $C_j$ and define $\chi^j(N_i) = d^i$ if $\text{IN}(N_i) = 1$ otherwise define $\chi^j(N_i) = -d^i$. Repeating this procedure until all $\chi^j(N_i), j = 0, \ldots, nv$ are obtained. Finally, it can be calculated that $\chi^s(N_i) = \max(\chi^1(N_i), \ldots, \chi^{nv}(N_i))$. Once the value of $\chi^s$ on every FE node is calculated, then one can carry out the X-FEM analysis following the same way as in [27, 28]. It is also worth noting that if structural components with more complex shapes are introduced in problem formulation, the more general approach developed in [28] can be employed to construct the corresponding TDFs.

### 4.2 Sensitivity analysis

For both the MMC and MMV-based approaches, under the assumption that the objective/constraint functional $I$ is differentiable with respect to design variables, the general form of the partial derivative of $I$ with respect to a design variable $d_i$ can be expressed as follow

$$\frac{\partial I}{\partial d_i} = \int_{\partial \Omega^s} f(\boldsymbol{u}, \boldsymbol{w}) \delta x_n^i \mathrm{d}S, \qquad (4.1)$$

where $\boldsymbol{u}$ and $\boldsymbol{w}$ are the so-called primary and adjoint fields, respectively. In Eq. (4.1), the integration is performed along the movable part of the boundary of the structure. When $I$ is the compliance of the structure we have $\boldsymbol{u} = \boldsymbol{w}$ and $f(\boldsymbol{u}, \boldsymbol{w}) = f(\boldsymbol{u}, \boldsymbol{w}) = -\mathbb{E}^s : \boldsymbol{\varepsilon}(\boldsymbol{u}) : \boldsymbol{\varepsilon}(\boldsymbol{u}) = -E_{ijkl}^s u_{i,j} u_{k,l}$ where $\mathbb{E}^s, \boldsymbol{u}$ and $\boldsymbol{\varepsilon}(\boldsymbol{u}) = \mathrm{sym}\nabla(\boldsymbol{u})$ denote the fourth-order elasticity tensor of the solid material, the displacement field and the strain tensor field, respectively. When $I$ represents the volume of the solid material, we have $f(\boldsymbol{u}, \boldsymbol{w}) = 1$. In Eq. (4.1), $\delta x_n^i$ denotes the variation of the boundary along the outward normal direction due to the variation of $d_i$ (i.e., $\delta d_i$). The expression of $\delta x_n^i$ can be found in the Appendix. We also refer the readers to [24, 27-29] for more details on calculating $\delta x_n^i$ from the variation of $d_i$ in the MMC and MMV-based approaches, respectively.

**4.3 Treatment of geometry constraints**

In the MMV-based approach, the constraint $\boldsymbol{n}^{c_0} \cdot \boldsymbol{b}_p \leq \cos\bar{\theta}, \forall \, \boldsymbol{x} \in C_0 \cap \mathrm{D}$ is actually pointwise in nature. In order to deal with this constraint, we discretize $C_0$ with a set of points $\boldsymbol{x}_j \in C_0 \cap \mathrm{D}, j = 1, \ldots, n$ and impose the constraints on these points (i.e., $g_i = \boldsymbol{n}_i^{c_0} \cdot \boldsymbol{b}_p - \cos\bar{\theta} \leq 0, i = 1, \ldots, n$). For the sake of reducing the computational effort, these constraints are aggregate into an equivalent single global constraint function as in [17]. Actually, $g_i = \boldsymbol{n}_i^{c_0} \cdot \boldsymbol{b}_p - \cos\bar{\theta} \leq 0, i = 1, \ldots, n$ is fully equivalent to

$$\mathcal{H}(g_1, \ldots, g_n) \leq 0, \qquad (4.2)$$

where

$$\mathcal{H}(g_1, \ldots, g_n) = \sum_{i=1}^n \begin{cases} g_i^2, & \text{if } g_i > 0, \\ 0, & \text{otherwise.} \end{cases} \qquad (4.3)$$

Compared to other approximation based approaches (e.g., P-norm approach and Kreisselmeier-Steinhauser (K-S) function approach), the advantage of this treatment is that there is no artificial parameter involved in this aggregation scheme and thus can make the original constraints being satisfied exactly.

Another type of constraints in the MMV-based approach is the non-intersection constraint for any two different printable features. It is require that $\Omega^{V_i} \cap \Omega^{V_j} = \emptyset, i \neq j, i,j = 1, \cdots, nv$ where $\Omega^{V_i}$ and $\Omega^{V_j}$ are the regions enclosed by $C_i$ and $C_j$, respectively. As first suggested in [31] and further refined in [32], the aforementioned non-intersection constraint can be represented in a single mathematical expression analytically as

$$\int_D H(\min(\chi^i(\boldsymbol{x}), \chi^j(\boldsymbol{x}))) \mathrm{d}V = 0, \tag{4.4}$$

where $\chi^i(\boldsymbol{x})$ and $\chi^j(\boldsymbol{x})$ denote the TDF of $\Omega^{V_i}$ and $\Omega^{V_j}$, respectively. In numerical implementation, $\chi^i(\boldsymbol{x})$ and $\chi^j(\boldsymbol{x})$ in Eq. (4.4) are replaced by $\chi^{i-\delta}(\boldsymbol{x})$ and $\chi^{j-\delta}(\boldsymbol{x})$ which represent the TDF of a δ-expension of $\Omega^{V_i}$ and $\Omega^{V_j}$, respectively (see Fig. 14 for a schematic illustration). This treatment can not only enhance the robustness of the solution process but also control the minimum length scale of the structure in an implicit way. In the present work, we take $\delta = \min(\Delta x, \Delta y)$ with $\Delta x$ and $\Delta y$ denoting the mesh size along two coordinate directions. Furthermore, the derivative of $\chi^i(\boldsymbol{x})$ with respect to design variables can be obtained by a finite difference scheme suggested in [25, 27].

For the MMV-based approach, the admissible set $\mathcal{U}_D$ that geometry design variables belongs to should also be determined appropriately. Actually in order to ensure the self-supporting property of the structure, we also require that

$$x_c^i \leq 0 \text{ if } \Omega^{V_i} \cap D_L \neq \emptyset \text{ and } x_c^i \geq L \text{ if } \Omega^{V_i} \cap D_R \neq \emptyset, \forall i = 1, \ldots, nv, \tag{4.5}$$

respectively. The meanings of the symbols appeared in the Eq. (4.5) can be understood from Fig. 15. It is also worth noting that the conditions $\Omega^{V_i} \cap D_L \neq \emptyset$ and $\Omega^{V_i} \cap D_R \neq \emptyset$ can also be expressed in the following analytical form

$$\int_{D \cup D_L \cup D_R} H(\min(\chi^i(\boldsymbol{x}), \chi^{D_L}(\boldsymbol{x}))) \mathrm{d}V > 0 \tag{4.6a}$$

and

$$\int_{D \cup D_L \cup D_R} H(\min(\chi^i(\boldsymbol{x}), \chi^{D_R}(\boldsymbol{x}))) \mathrm{d}V > 0, \tag{4.6b}$$

respectively. In Eq. (4.6), $\chi^i(\boldsymbol{x}), \chi^{D_L}(\boldsymbol{x})$ and $\chi^{D_R}(\boldsymbol{x})$ denote the TDF of $\Omega^{V_i}$, $D_L$ and $D_R$, respectively.

In the present work, the exterior boundary of the structure is also represented by a B-spline curve (i.e., $C_0$) shown in Fig. 16 where 10 control points are adopted. The only constraint imposed on the coordinates of these control points is $y_5 \leq -R$, where $R$ is a sufficiently large positive number. This treatment guarantees that the exterior boundary cannot be of the wavy shape and is totally separated from the baseplate since these two situations will inevitably lead to the existence of unprintable structures (see Fig. 16 for reference).

## 5. Numerical examples

In this section, several numerical examples are provided to demonstrate the effectiveness of the proposed approaches for designing self-supporting structures through topology optimization. Since the main purpose of the present section is to test the numerical performance of the suggested approach, the material, load and geometric data are all chosen as dimensionless unless otherwise stated. The Young's modulus and Poisson's ratio of the solid material are taken as $E = 1$ and $\nu = 0.3$, respectively. Plane stress state (with unit thickness) is assumed and four-node bilinear square elements are adopted for finite element analysis in all presented examples. The Method of Moving Asymptotes (MMA) [33] is adopted as numerical optimizer to solve the optimization problems. Structural compliance is taken as the objective functional and the printability constraints as well as volume constraint on available solid material are always considered in all examples. For all examples, B-splines with six control points are adopted to describe the shape of each printable void (i.e., $n = 6$ in Eq. (2.8) and among of them only five control points are independent) and B-splines with ten control points are used to characterize the exterior boundary of the structures, respectively, in the MMV-based approach. Furthermore, the lower bound of the overhang angle is set to $\bar{\theta} = 45°$.

### 5.1 Tensile beam example

A simple tensile beam example [21] is first considered in order to test the effectiveness of the proposed approaches. In this example, a rectangular design domain which is discretized by a $160 \times 80$ FE mesh is shown in Fig. 17. The left side of the design domain is fixed and distributed horizontal loads with magnitude of 1 are imposed on the right top side of the design domain. It is assumed that the print direction is $\boldsymbol{b}_p = (0,1)^\top$.

Firstly, this example is tested by employing MMC-based approach formulated in Eq. (2.1). The initial design composed of 16 components is shown in Fig. 18 and the initial rotation angles (i.e., $\alpha$) of the working plane is set to $45°$. It is assumed that the upper bound of the solid material is $\overline{V} = 20\%$. The optimized structure is shown in Fig. 19, which is almost the same as that without considering the self-supporting constraint (a horizontal strait beam). The final optimized rotation angles of the working plane is $\alpha = 75.93°$. Here the total number of design variables equals $16 \times 5 + 1 = 81$. Some intermediate optimization results obtained during the optimization process is given in Fig. 20.

The problem is also solved by the MMV-based approach formulated in Eq. (2.9). This time, the upper bound of the relative available volume constraint is set to $\overline{V} = 40\%$ since part of solid material will be used to form support structure. Two different initial designs containing 10 and 15 non-overlap printable voids (shown in Fig. 21a and Fig. 21b, respectively) are considered. There are totally $10 \times (4 + 2) + 10 = 70$ (i.e., 4 independent control points for each printable void and 1 center point as well as 10 control points for the exterior structural boundary) and $15 \times (4 + 2) + 10 = 100$ design variables for these two initial designs, respectively. Optimized structures obtained from the two initial designs are shown in Fig. 22, respectively. It can be seen from these figures that besides a horizontal solid beam at the top side of the structure which constitutes the main load transition path, some of the solid material (about 15%) has been used to form the supporting structure between the horizontal beam and baseplate in the optimized designs. These two optimized designs are obviously satisfied the self-supporting requirement since every part of the structures is sufficiently supported along print direction and there is actually no component whose inclined angle is larger than the prescribed critical value (i.e., $\overline{\theta} = 45°$). Numerical results also indicate that as the number of printable voids included in the initial design increases, the performance (measure in terms of the value of the objective functional) of the structure improves. This is quite consistent with our theoretical analysis made in Section 3 which states that the optimal support structure may be constructed by many thin components forming a hierarchical network. It is also obvious that the creation of the support structure is purely to satisfy the printability requirement. Furthermore, compared to the optimized designs show in [21], the optimized designs obtained by the proposed approach are totally free from the existence of undesirable grey elements.

Fig. 23 shows some intermediate optimization results obtained during the course of optimization where the initial design shown in Fig. 21b is adopted. From this figure it can be observed clearly how printable topologies are reached by changing the positions and shapes of the voids. Corresponding convergence iteration histories are also provided in Fig. 24. It can be observed from this figure (and corresponding figures in the following examples), in the MMV approach, the objective function can be reduced to a stable value within 100 iterations and the subsequent iteration steps are used to satisfy the AM related geometry constraints.

**5.2 Short cantilever beam example**

In this example, the well-known short beam problem is examined. The design domain, external load, and boundary conditions are all shown in Fig. 25. A rectangular design domain has the width of $W = 2$ and length $L = 1$ is discretized by a $120 \times 60$ FE mesh. A unit vertical load is imposed on the middle point of the right side of the design domain. For this problem it is assumed that $\bar{V} = 50\%$ and $\boldsymbol{b}_p = (0,1)^\top$.

Firstly, the optimized structure obtained without considering the self-supporting constraint is shown in Fig. 26. The corresponding optimal value of the objective functional is $I = 61.59$. This structure is, however, not self-supportive since the overhang angles of some structural parts are obviously less than the prescribed critical overhang angle $\bar{\theta} = 45°$.

Secondly, the problem is solved by employing MMC-based approach formulated in Eq. (2.1). The initial designs shown in Fig. 27a and Fig. 27b are composed of 16 components and the initial rotation angles of the working plane are set to $45°$ and $90°$, respectively. The total number of design variables is actually only $16 \times 5 + 1 = 81$ in this MMC-based approach. Corresponding optimization results are shown in Fig. 28a and Fig. 28b, respectively. It can be observed from these figures that the proposed approach does have the capability to find the optimal value of the rotation angle of the working plane. The horizontal components in Fig. 26 which are very effective to transfer the external load but unprintable when $\alpha = 0°$, can now be printed by rotating the working plane to $\alpha = 90°$. This demonstrates the advantage of adopting the rotation angle as one of the design variables in the problem formulation. Some intermediate steps of the optimization process starting from the initial design shown in Fig. 27a are shown in Fig. 29. It can be seen from this figure that compared to the case where no self-supporting requirement is considered and $\alpha = 0°$ is fixed during the process of optimization, the symmetry property of the problem is lost and the final optimized structure is not symmetric anymore. It is also worth noting that the optimal value of the objective functional is $I = 62.8$ which is very close to the value of the structure shown in Fig. 26. This demonstrates once again the necessity of introducing the rotation angle of the working plane as a design variable. In addition, if we start the optimization from $\alpha = 90°$, it can be observed from Fig. 30 that the optimized value of the rotation angle of the working plane keeps the same value. The value of the objective functional for this structure is $I = 62.55$. The values of the inclined angles of some components in the optimized structure shown in Fig. 28 are listed in Table 1. Corresponding iteration history is also plotted in Fig. 31. Compared with the MMV approach, convergence is more rapid in the MMC approach since less number of geometry constraints are included in the problem formulation. For this example, optimized structure is obtained within 300 iterations.

Lastly, the considered problem is solved with the MMV-based approach formulated in Eq. (2.9). As shown in Fig. 32 (where $\boldsymbol{b}_p = (0,1)^\top$) and Fig. 33 ($\boldsymbol{b}_p = (1,0)^\top$), twelve printable voids are distributed in the design domain as the initial design. Under this circumstance, the total number of design variables is $12 \times (4 + 2) + 10 = 82$. Fig. 34 and Fig. 35 plot the optimized structures obtained under different print directions. It can be observed that the self-supporting requirement does have been satisfied by these structures. The obtained structures are totally black-and-white and have crisp boundaries. These features cannot be achieved easily with use of traditional methods. It is also worth noting that the optimal structural topology is highly dependent on the print direction when self-supporting requirement is considered. When $\boldsymbol{b}_p = (0,1)^\top$, the corresponding optimized structure is not symmetric and the value of the objective functional for this structure is $I = 72.164$. This non-symmetric behavior is also consistent with the observation made in [23]. If, however, the print direction is changed to $\boldsymbol{b}_p = (1,0)^\top$, the obtained optimized structure is very similar to the one shown in Fig. 26 and the compliance of this structure is $I = 63.408$ which is also very close to the value associated with the structure obtained without considering the self-support constraint (i.e., $I =$

61.59).

Fig. 36 provides some intermediate results during the process of optimization where $\boldsymbol{b}_p = (1,0)^\top$. It can be observed from this figure that although the voids are restricted to be non-overlap, significant topology changes can still be achieved during the course of optimization. Twelve voids in the initial design finally reduce to only three ones through shrinking or moving outside. This demonstrates clearly the capability of the MMV-based approach to deal with topology changes. Fig. 37 plots the histories of the values of the objective functional and constraint functions during the process of numerical optimization.

**5.3 MBB-beam example**

In this example, a MBB example will be investigated. The setting of this problem is described schematically in Fig. 38. A vertical load is imposed on the middle point of the top side of a rectangular design domain，which is discretized by a $360 \times 60$ FE mesh. The upper bound on the relative available volume of the solid material (for half of the structure) is $\bar{V} = 50\%$.

Fig. 39 plots the optimization results obtained without considering self-supporting requirement for comparison purpose. The corresponding value of the objective functional is $I = 381.80$. Since there exist several horizontal structural members in this structure, it is not printable when $\boldsymbol{b}_p = (0,1)^\top$.

Next, the same problem is solved by employing the MMC-based approach formulated in Eq. (2.1) where the layout of the structural components and the rotational angle of the working plane are all taken as design variables. The initial design shown in Fig. 40 contains 48 straight components (the corresponding number of design variables is 121) and the initial rotation angle of the working plane is set to $\alpha = 45°$. The optimized structure is shown in Fig. 41 and the corresponding value of the objective functional is $I = 386.36$. Besides, the optimized value of the rotation angle is $\alpha = 84.87°$. For this optimized solution, the compliance is only increased 1% compared to the solution without considering the self-supporting requirement ($I = 381.80$). This demonstrates clearly that it is necessary to include the print direction as a design variable in AM oriented topology optimization problems.

Finally, the MMV-based approach formulated in Eq. (2.9) is also used to solve this problem.

Since the problem under consideration is symmetric in nature, only half of the structure is considered. As shown Fig. 42, there are totally 14 printable voids in the initial design and the total number of design variables is 94. Fig. 43a plots the optimized design where printable voids with different sizes are distributed separately to achieve high structural efficiency. The corresponding compliance is $I = 436.08$, which is only approximately 14% higher than that of the unconstrained optimal solution ($I = 381.80$). This is of course the price that should be paid for considering the self-supporting constraint. The obtained structure is obviously self-supporting and quite reasonable from mechanics point of view. From Fig. 43b, it is also noteworthy that the optimized structure inherits most the optimal feature of the unconstrained design, for example, three printable voids (but now with more large inclined angles) and a relatively long horizontal top chord. It is also interesting to observe that two unprintable voids in the unconstrained structure have been replaced by six printable voids (which also provide sufficient support on the top chord) in the optimized self-supporting structure. In some sense, this optimality mechanism is also consistent with the theoretical analysis made in Section 3. Moreover, the explicit piecewise expression of the $\boldsymbol{C}_1$ boundary curve in Fig. 43a is: the first segment:

$$\boldsymbol{C}_1(u) = \begin{Bmatrix} x \\ y \end{Bmatrix} = \begin{Bmatrix} -10.72u^2 + 4.76u + 2.05 \\ 7.29u^2 - 4.89u + 0.87 \end{Bmatrix}, \quad 0 \leq u \leq 0.33, \quad (5.1a)$$

the second segment:

$$\boldsymbol{C}_1(u) = \begin{Bmatrix} x \\ y \end{Bmatrix} = \begin{Bmatrix} 0.09u^2 - 2.44u + 3.25 \\ 0.07u^2 - 0.07u + 0.07 \end{Bmatrix}, \quad 0.33 \leq u \leq 0.66 \quad (5.1b)$$

and the third segment

$$\boldsymbol{C}_1(u) = \begin{Bmatrix} x \\ y \end{Bmatrix} = \begin{Bmatrix} 10.45u^2 - 16.25u + 7.85 \\ 7.29u^2 - 9.70u + 3.28 \end{Bmatrix}, \quad 0.66 \leq u \leq 1, \quad (5.1c)$$

respectively. Fig. 44 plots some intermediate optimization results.

## 6. Concluding remarks

In the present paper, AM oriented topology optimization methods that can generate self-supporting structures are developed under the MMC and MMV solution frameworks. Numerical examples show that the proposed approaches (especially the MMV-based approach) do have the capability of finding optimized designs where overhang angle constraints can be fully respected.

Besides optimal structural topology, the build orientation of AM can also be optimized with use of the proposed approaches in a straightforward way. Compared with existing approaches, the distinctive feature of the present approach is that it solves the corresponding problem through a more explicit and geometrical treatment. As for other AM related manufacture constraints, it is worth noting that as shown in [17], the proposed MMC-based approach also has potential to deal with minimum length scale constraint. Furthermore, for two dimensional problems the issue of enclosed voids can be dealt with easily using the proposed MMV-based approach by imposing the constraint such that there is no printable features existing in the interior of a design domain. Extending the proposed methods to three dimensional (3D) problems can be achieved by introducing some 3D printable features like the one shown in Fig. 45. Corresponding results will be reported elsewhere.

**Appendix**

In this appendix, sensitivity analysis associated with the MMV approach will be outlined briefly. Actually, with use of the B-spline curve description scheme, the variation of the boundary corresponding to $C_i(u)$ along the outward normal direction in Eq. (4.1) can be calculated as:

$$\delta x_n^i = \delta C_i(u) \cdot n_i, \tag{A1}$$

where $n_i$ is the outward normal vector of the boundary and it can be calculated from the relations

$$t_i \cdot n_i = 0 \tag{A2a}$$

and

$$\|n_i\| = 1. \tag{A2b}$$

In Eq. (A2a),

$$t_i = \frac{dC_i(u)}{du} = \sum_{k=0}^{n} \frac{dN_{k,p}(u)}{du} P_k^i, \quad a \le u \le b. \tag{A3}$$

The expression of $\delta C_i(u)$ can be calculated from Eq. (2.8a) in the main text as

$$\delta C_i(u) = \sum_{k=0}^{n} N_{k,p}(u) \delta P_k^i, \quad a \le u \le b, \tag{A4}$$

with

$$\delta P_k^i = \left(\delta x_c^i + \sin\left((k-1)\theta_k^i + \frac{\pi}{2}\right)\delta d_k^i, \delta y_c^i + \cos\left((k-1)\theta_k^i + \frac{\pi}{2}\right)\delta d_k^i\right)^\mathsf{T},$$

$$\theta = \frac{\pi}{n-2}, \quad k = 1, \ldots, n-1, \tag{A5a}$$

and

$$\delta P_0^i = \delta P_n^i = \left(\delta x_c^i, \delta y_c^i + \delta d_0^i\right)^\mathsf{T}. \tag{A5b}$$

The above derivation immediately leads to the conclusion that $\delta x_n^i = \delta C_i(u) \cdot n_i = \sum_{j=0}^{n}(A_j^i \delta x_c^i + B_j^i \delta y_c^i + C_j^i \delta d_j^i)$. The detailed expressions of $A_j^i, B_j^i$ and $C_j^i$ are omitted here since the corresponding derivation is a trivial task.

*Computer Methods in Applied Mechanics and Engineering, under review*  2016-11-12  **25**

**Acknowledgements**

The financial supports from the National Key Research and Development Plan (2016YFB0201600), the National Natural Science Foundation (11402048, 11372004), Program for Changjiang Scholars, Innovative Research Team in University (PCSIRT) and 111 Project (B14013) are gratefully acknowledged.

# References

[1] K.T. Cheng, N. Olhoff, An investigation concerning optimal design of solid elastic plate, Int. J. Solids Struct. 17 (1981) 305–323.

[2] M.P. Bendsøe, N. Kikuchi, Generating optimal topologies in structural design using a homogenization method, Comput. Methods Appl. Mech. Engrg. 71 (1988) 197–224.

[3] G.I.N. Rozvany, A critical review of established methods of structural topology optimization, Struct. Multidiscip. Optim. 37 (2009) 217–237.

[4] S. Nelaturi, W. Kim, T. Kurtoglu, Manufacturability feedback and model correction for additive manufacturing, ASME J. Manuf. Sci. Engrg. 137 (2015) 21015-1-021015-9.

[5] R. Mertens, S. Clijsters, K. Kempen, J.P. Kruth, Optimization of scan strategies in selective laser melting of aluminum parts with downfacing areas, ASME J. Manuf. Sci. Engrg. 136 (2014) 61012-1-061012-7.

[6] J. Krantz, D. Herzog, C. Emmelmann, Design guidelines for laser additive manufacturing of light weight structures in TiAl6V4, J. Laser Appl. 27 (2015) S14001-1- S14001-16.

[7] D. Brackett, I. Ashcroft, R. Hague, Topology optimization for additive manufacturing, Proceedings of the 24th Solid Freeform Fabrication Symposium, Austin, TX, 2011, 348–362.

[8] J.K. Liu, Y.S. Ma, A survey of manufacturing oriented topology optimization methods, Adv. Eng. Softw. 100 (2016) 161-175.

[9] T.A. Poulsen, A new scheme for imposing minimum length scale in topology optimization, Int. J. Numer. Methods Engrg. 57 (2003) 741–760.

[10] J.K. Guest, Imposing maximum length scale in topology optimization, Struct. Multidiscip. Optim. 37 (2009) 463–473.

[11] S.K. Chen, M.Y. Wang, A.Q. Liu, Shape feature control in structural topology optimization, Comput. Aided Des. 40 (2008) 951–962.

[12] X. Guo, W.S. Zhang, W.L. Zhong, Explicit feature control in structural topology optimization via level set method, Comput. Methods Appl. Mech. Engrg. 272 (2014) 354–378.

[13] W.S. Zhang, W.L. Zhong, X. Guo, An explicit length scale control approach in SIMP-based topology optimization, Comput. Methods Appl. Mech. Engrg. 282 (2014) 71–86.

[14] Q. Xia, T.L. Shi, Constraints of distance from boundary to skeleton: For the control of length scale in level set based structural topology optimization, Comput. Methods Appl. Mech. Engrg. 295 (2015) 525-542.

[15] M.D. Zhou, B.S. Lazarov, F.W. Wang, O. Sigmund, Minimum length scale in topology optimization by geometric constraints, Comput. Methods Appl. Mech. Engrg. 293 (2015) 266-282.

[16] G. Allaire, F. Jouve, G. Michailidis, Thickness control in structural optimization via a level set method, Struct. Multidiscip. Optim. 53 (2016) 1349–1382.

[17] W.S. Zhang, D. Li, J. Zhang, X. Guo, Minimum length scale control in structural topology optimization based on the Moving Morphable Components (MMC) approach, Comput. Methods Appl. Mech. Engrg. 311 (2016) 327–355.

[18] Q.H. Li, W.J. Chen, S. Liu, L.Y. Tong, Structural topology optimization considering connectivity constraint, Struct. Multidiscip. Optim. 54 (2016) 971-984.

[19] M. Leary, L. Merli, F. Torti, M. Mazur, M. Brandt, Optimal topology for additive manufacture: A method enabling additive manufacture of support-free optimal structures, Mater. Des. 63 (2014) 678-690.

[20] K. Hu, S. Jin, C.C. Wang, Support slimming for single material based additive manufacturing, Comput. Aided Des. 65 (2015) 1-10.

[21] M. Langelaar, An additive manufacturing filter for topology optimization of print-ready designs, Struct. Multidiscip. Optim. doi:10.1007/s00158-016-1522-2.

[22] M. Langelaar, Topology optimization of 3D self-supporting structures for additive manufacturing, Addit. Manuf. 12 (2016) 60-70.

[23] A.T. Gaynor, J.K. Guest, Topology optimization considering overhang constraints: Eliminating sacrificial support material in additive manufacturing through design, Struct. Multidiscip. Optim. doi:10.1007/s00158-016-1551-x.

[24] X. Guo, W.S. Zhang, W.L. Zhong, Doing topology optimization explicitly and geometrically—a new moving morphable components based framework, ASME Trans. J. Appl. Mech. 81 (2014) 081009-1–081009-12.

[25] W.S. Zhang, J. Yuan, J. Zhang, X. Guo, A new topology optimization approach based on Moving

Morphable Components (MMC) and the ersatz material model, Struct. Multidiscip. Optim. 53 (2016) 1243–1260.

[26] J. A. Norato, B.K. Bell, D.A. Tortorelli, A geometry projection method for continuum-based topology optimization with discrete elements, Comput. Methods Appl. Mech. Engrg. 293 (2015) 306–327.

[27] W.S. Zhang, J. Zhang, X. Guo, Lagrangian description based topology optimization—A revival of shape optimization, ASME Trans. J. Appl. Mech. 83 (2016) 041010-1–041010-16.

[28] X. Guo, W.S. Zhang, J. Zhang, Explicit structural topology optimization based on moving morphable components (MMC) with curved skeletons, Comput. Methods Appl. Mech. Engrg. 310 (2016) 711-748.

[29] W.S. Zhang, W.Y. Yang, J.H. Zhou, D. Li, X. Guo, Structural optimization through explicit boundary evolution, ASME Trans. J. Appl. Mech. 84 (2017) 011011-1–011011-10.

[30] P. Wei, M.Y. Wang, X.H. Xing, A study on X-FEM in continuum structural optimization using a level set model, Comput. Aided Des. 42 (2010) 708-719.

[31] P. Shan, Optimal embedding objects in the topology design of structure (Master thesis) Dalian University of Technology (2008) http://d.wanfangdata.com.cn/Thesis_Y1247462.aspx

[32] W.S. Zhang, W.Z. Zhong, X. Guo, Explicit layout control in optimal design of structural systems with multiple embedding components, Comput. Methods Appl. Mech. Engrg. 290 (2015) 290-313.

[33] K. Svanberg, The method of moving asymptotes—a new method for structural optimization, Int. J. Numer. Methods Engrg. 24 (1987) 359–373.

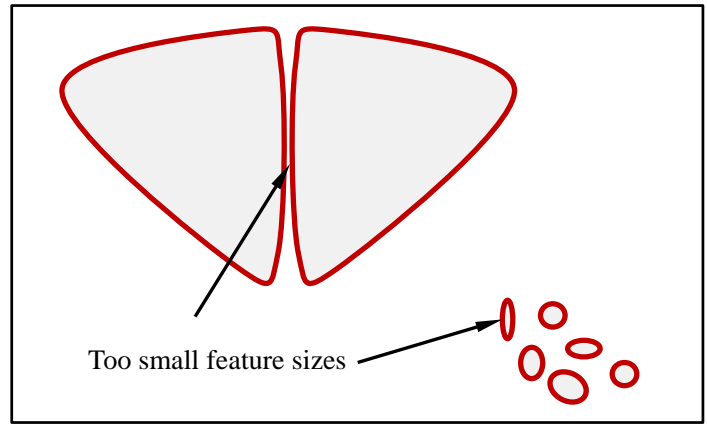

(a). Unprintable structural features.

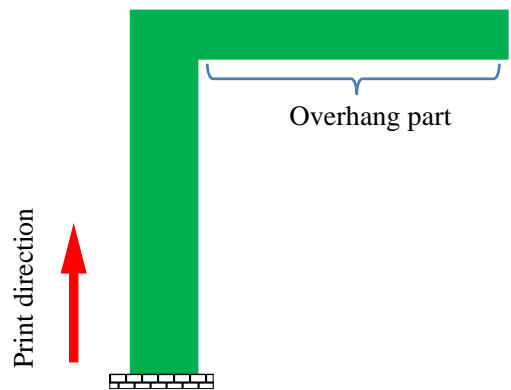

(b). An unprintable structural part without sufficient support from below.

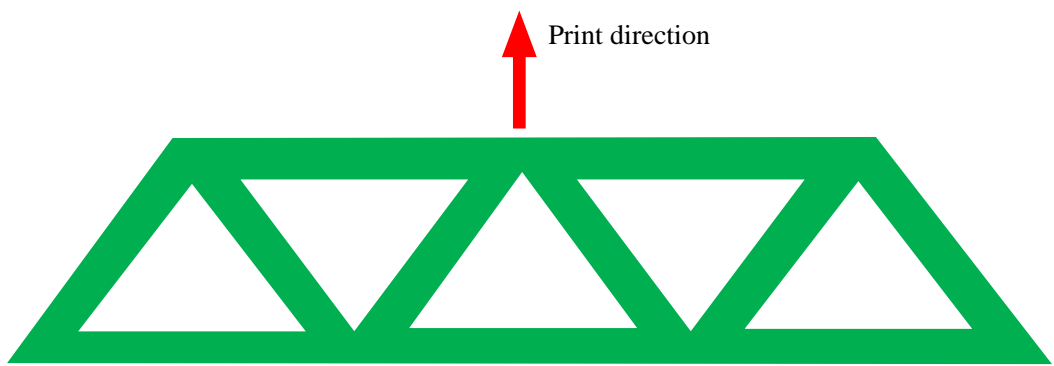

(c). A non-self-supporting structure.

Fig. 1 A schematic illustration of the design limitations in AM.

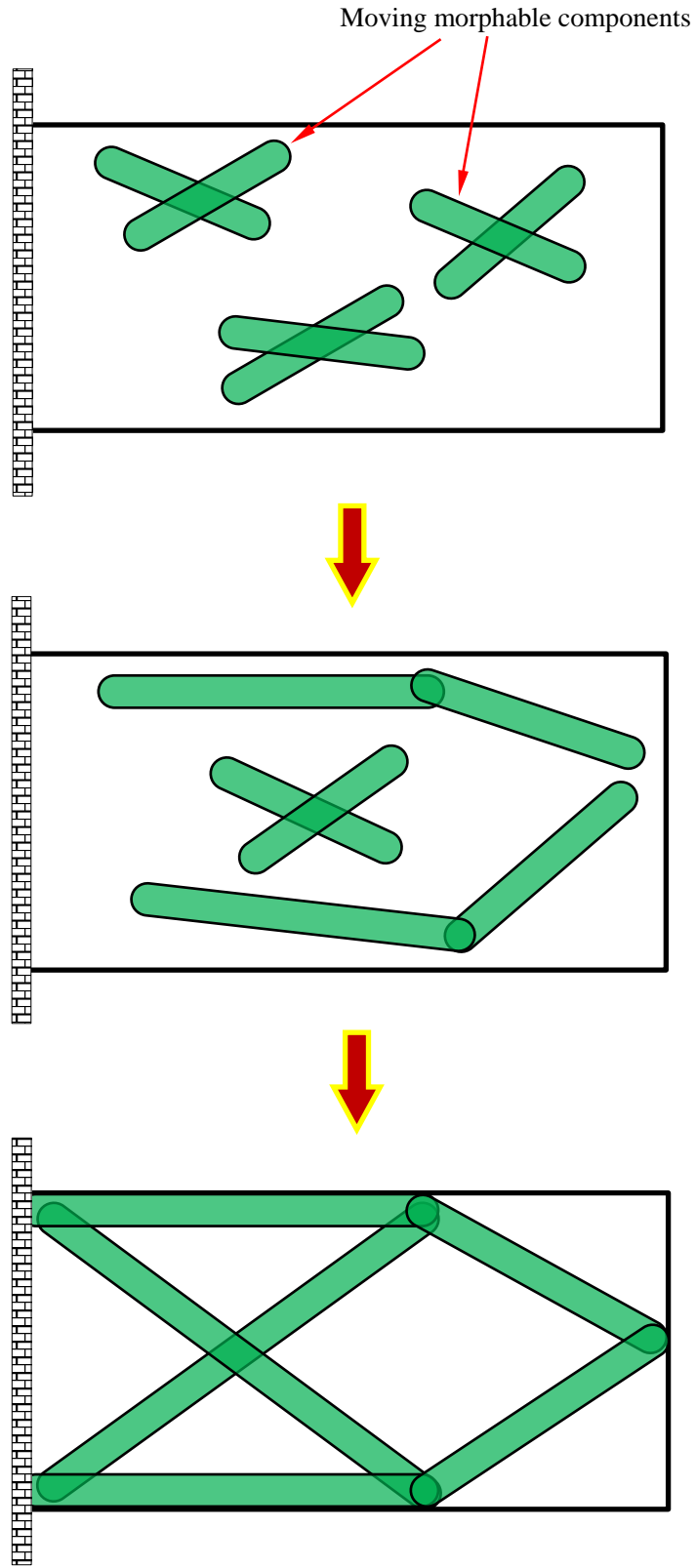

Fig. 2 The basic idea of the MMC-based topology optimization approach.

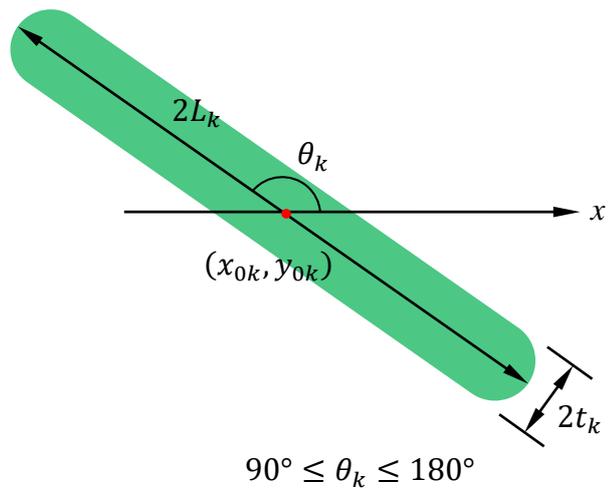

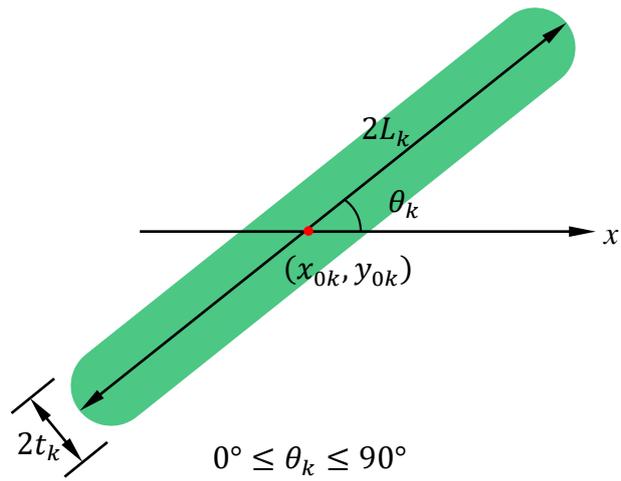

Fig. 3 Geometry description of a structural component in the MMC approach.

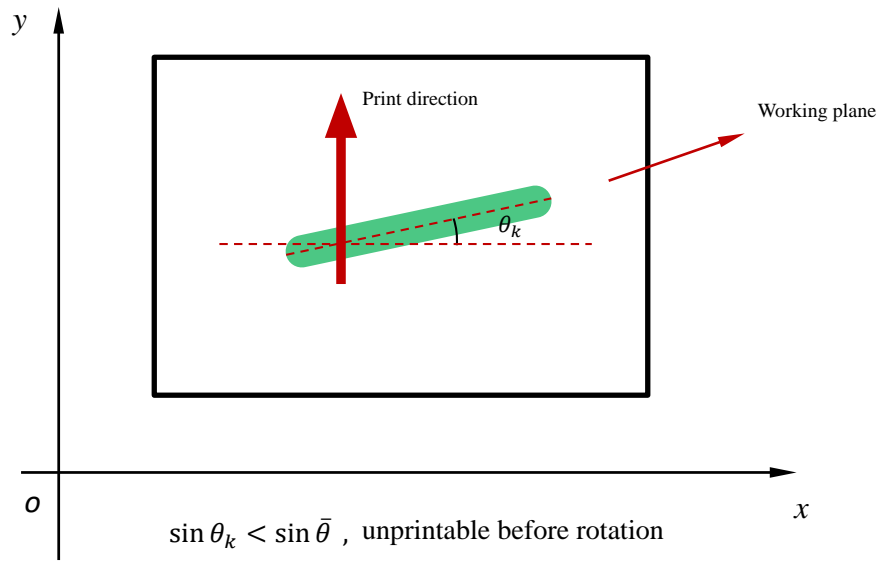

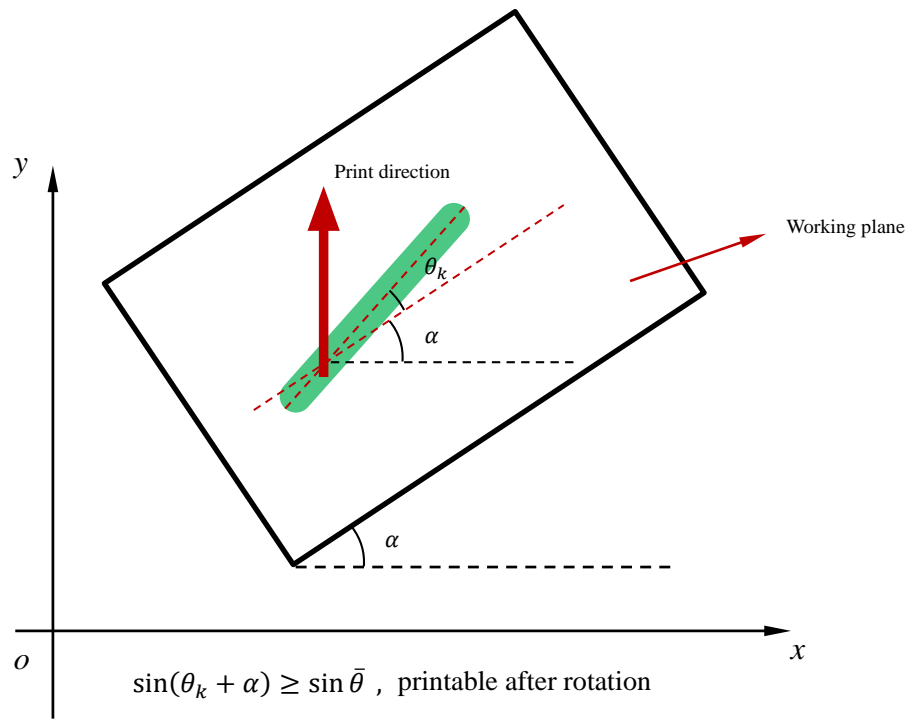

Fig. 4 The rotation angle of the working plane.

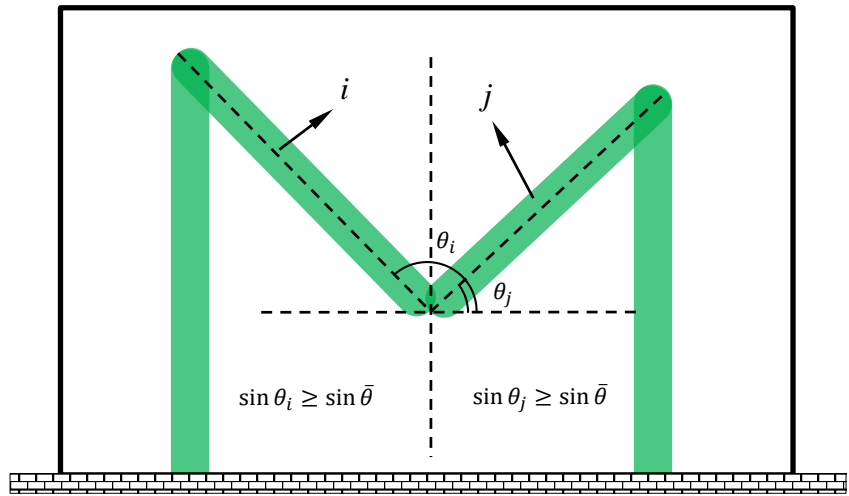

(a). An unprintable V-shape material distribution case.

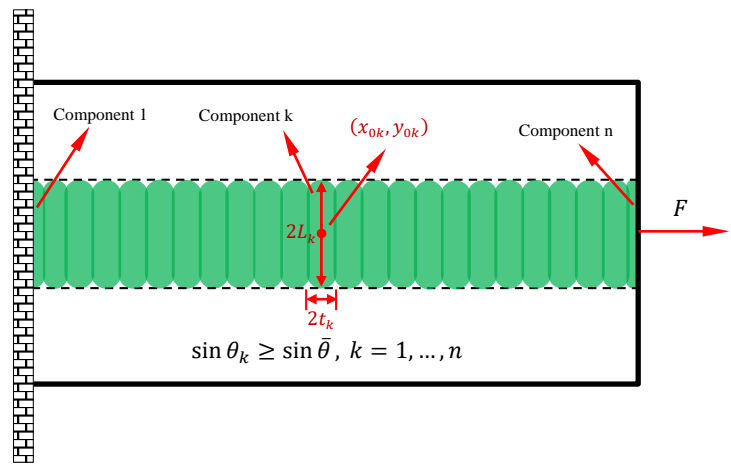

(b). Another unprintable components distribution case.

Fig. 5 An illustration of unprintable components distribution cases.

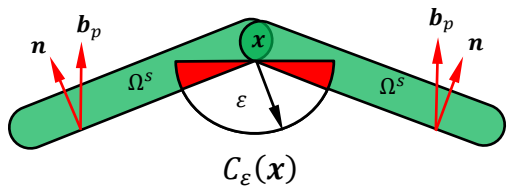
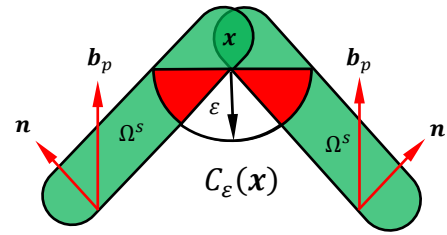

An unprintable case

$\text{meas}(\Omega^s \cap C_\varepsilon(\boldsymbol{x})) > 0, \quad \forall \boldsymbol{x} \in \partial\Omega^s$

$\boldsymbol{n} \cdot \boldsymbol{b}_p \geq \cos\bar{\theta}, \quad \forall \boldsymbol{x} \in \partial\Omega^s$

A printable case

$\text{meas}(\Omega^s \cap C_\varepsilon(\boldsymbol{x})) > 0, \quad \forall \boldsymbol{x} \in \partial\Omega^s$

$\boldsymbol{n} \cdot \boldsymbol{b}_p \leq \cos\bar{\theta}, \quad \forall \boldsymbol{x} \in \partial\Omega^s$

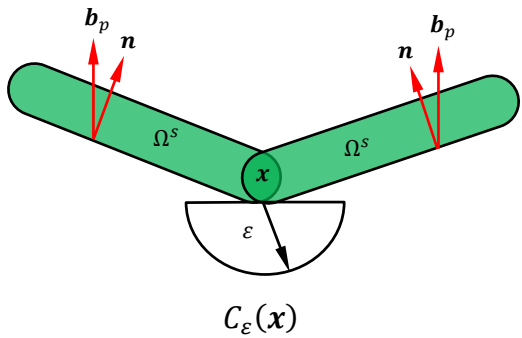
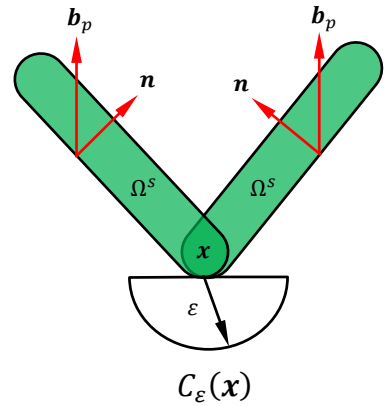

An unprintable case

$\text{meas}(\Omega^s \cap C_\varepsilon(\boldsymbol{x})) \leq 0, \quad \forall \boldsymbol{x} \in \partial\Omega^s$

$\boldsymbol{n} \cdot \boldsymbol{b}_p \geq \cos\bar{\theta}, \quad \forall \boldsymbol{x} \in \partial\Omega^s$

An unprintable case

$\text{meas}(\Omega^s \cap C_\varepsilon(\boldsymbol{x})) \leq 0, \quad \forall \boldsymbol{x} \in \partial\Omega^s$

$\boldsymbol{n} \cdot \boldsymbol{b}_p \leq \cos\bar{\theta}, \quad \forall \boldsymbol{x} \in \partial\Omega^s$

Fig. 6 A schematic illustration of the meanings of the constraints in Eq. (2.7a) and Eq. (2.7b).

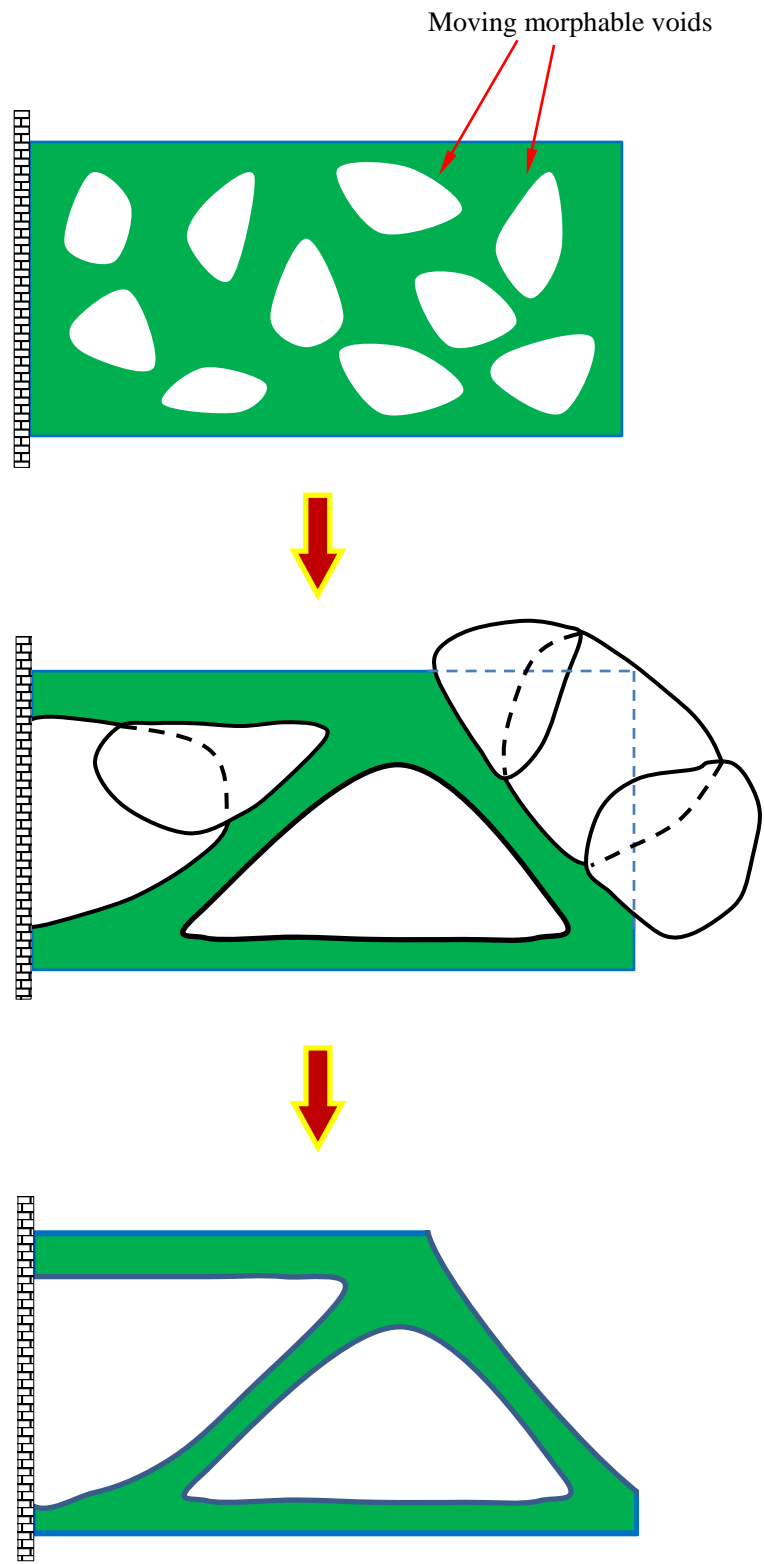

Fig. 7 The basic idea of the MMV-based topology optimization approach.

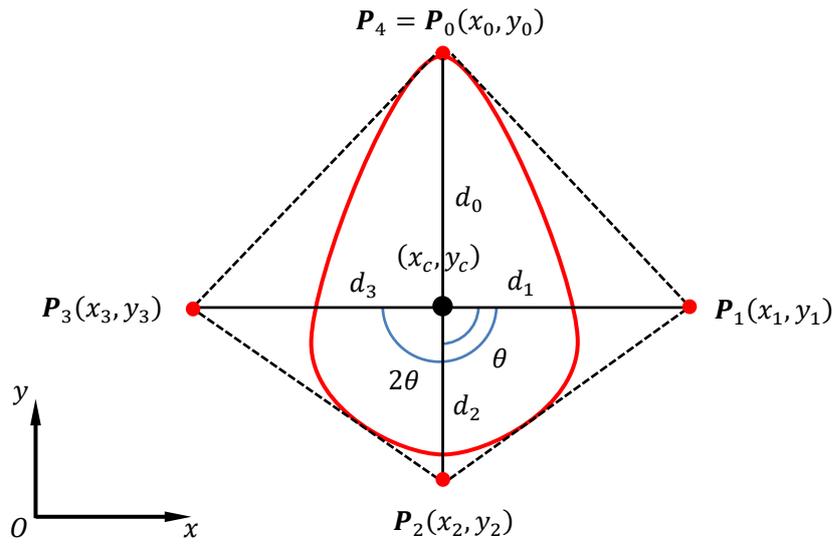

Fig. 8 The construction of a typical B-spline curve $C(u)$ described in Eq. (2.8).

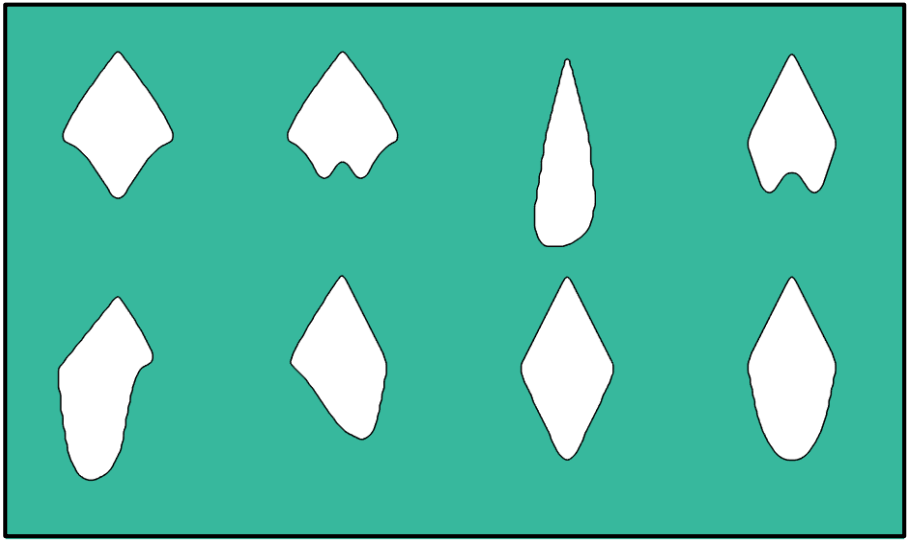

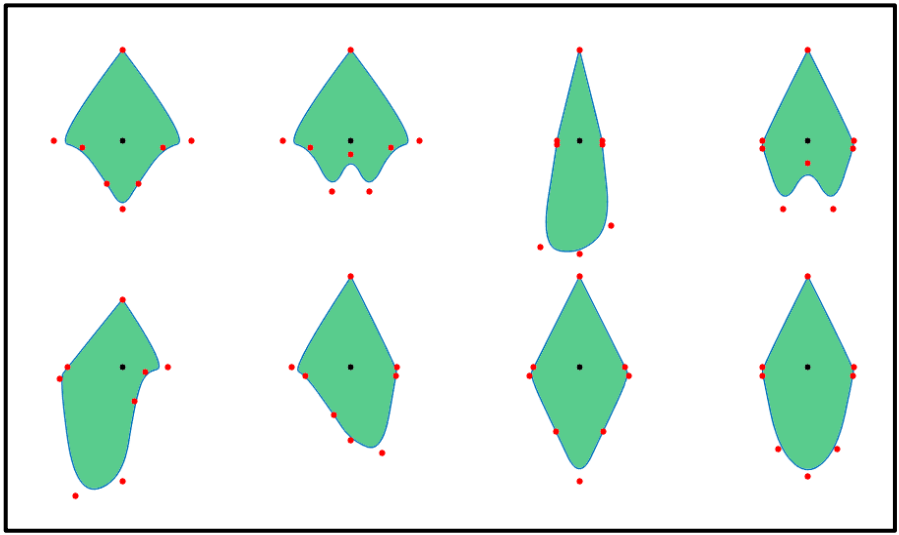

Fig. 9 Some printable features with complex shapes.

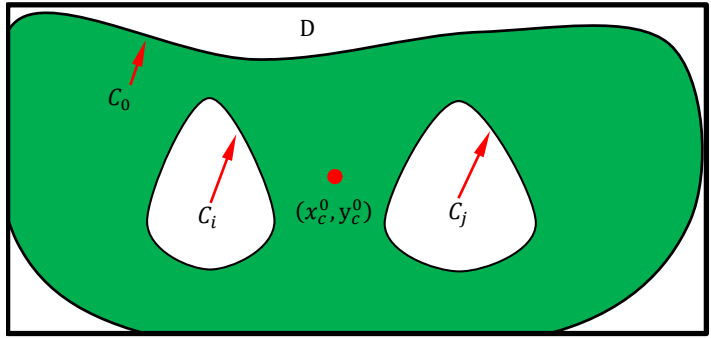

(a). $\Omega^{V_i} \cap \Omega^{V_j} = \emptyset$, $(\Omega^{V_i} \cup \Omega^{V_j}) \cap C_0 = \emptyset$ (printable case).

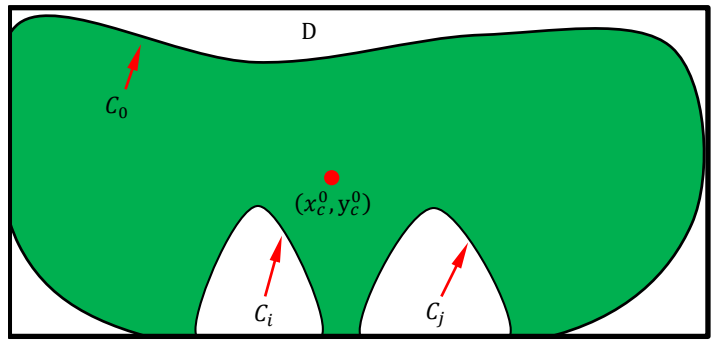

(b). $\Omega^{V_i} \cap \Omega^{V_j} = \emptyset$, $\Omega^{V_i} \cap C_0 \neq \emptyset$, $\Omega^{V_j} \cap C_0 \neq \emptyset$ (printable case).

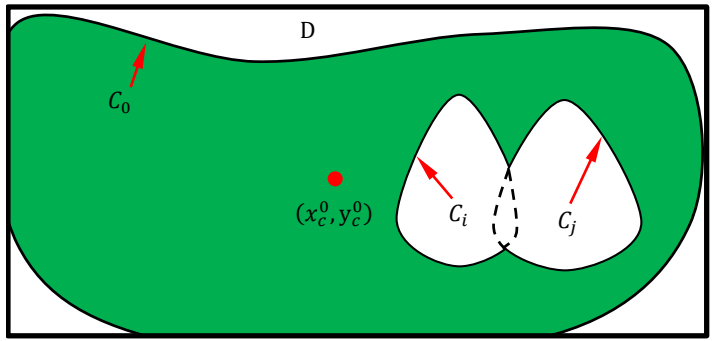

(c). $\Omega^{V_i} \cap \Omega^{V_j} \neq \emptyset$ (unprintable case).

Fig. 10 Printable and unprintable cases in the MMV-based approach.

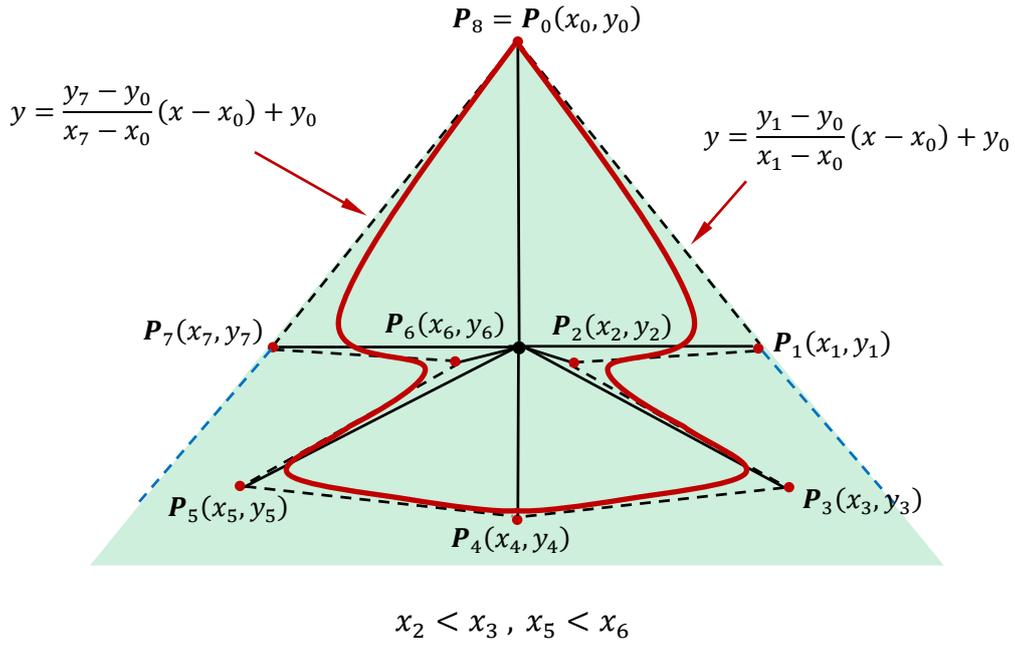

(a). An unprintable case.

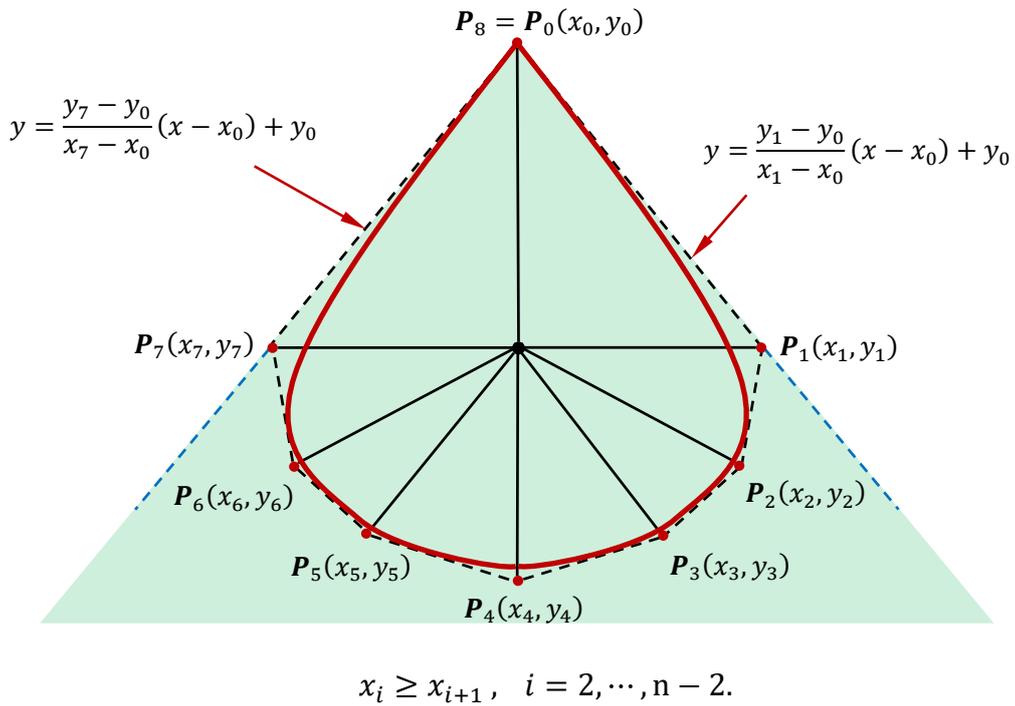

(b). A printable case.

Fig. 11 A schematic illustration of the construction of a printable void.

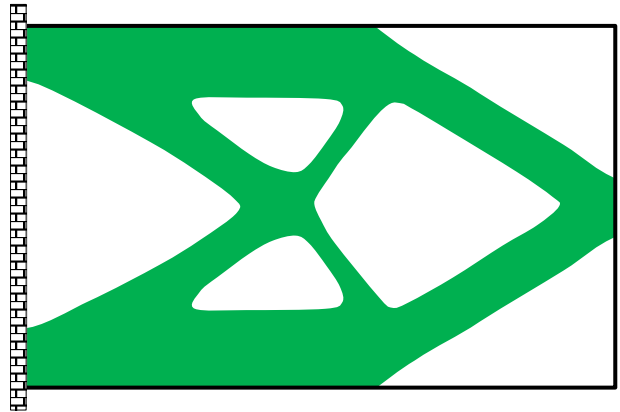

(a). Optimal structure without considering self-supporting requirement.

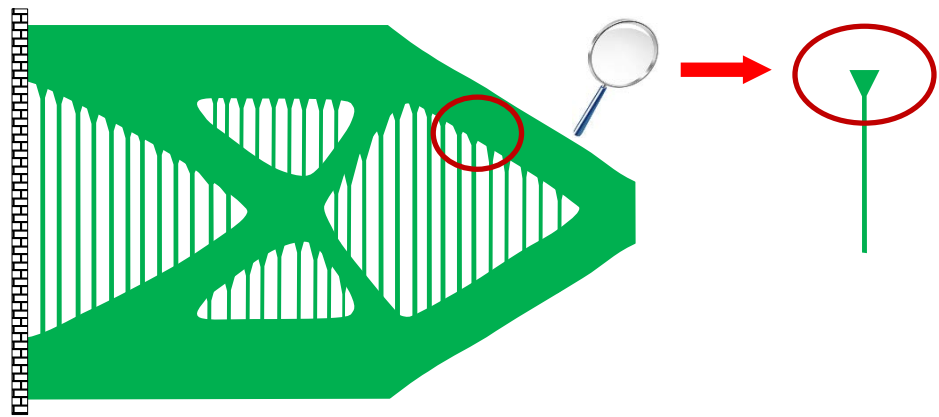

(b). A conjectured optimal self-supporting structure.

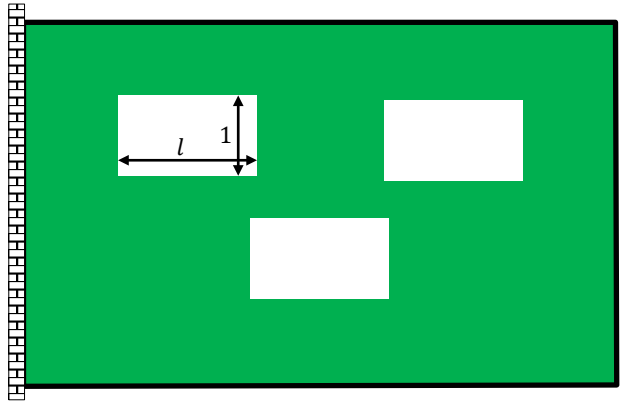

(c). An assumed non-self-supporting optimal structure with rectangular holes.

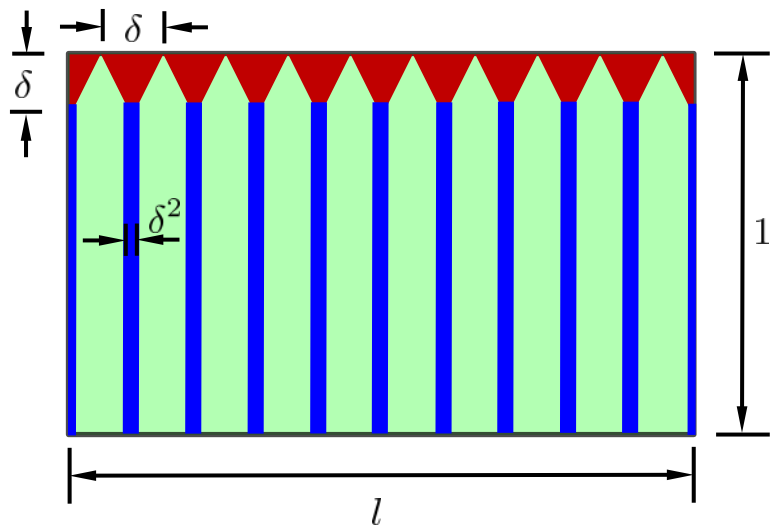

(d). Making the structure shown in Fig. (12c) self-supporting by adding support material.

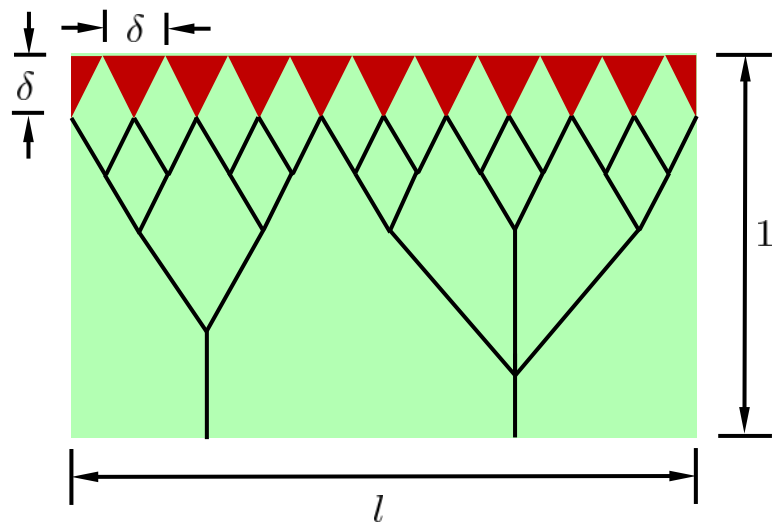

(e). A hierarchical support structure.

Fig. 12 Construction of an optimal self-supporting structure.

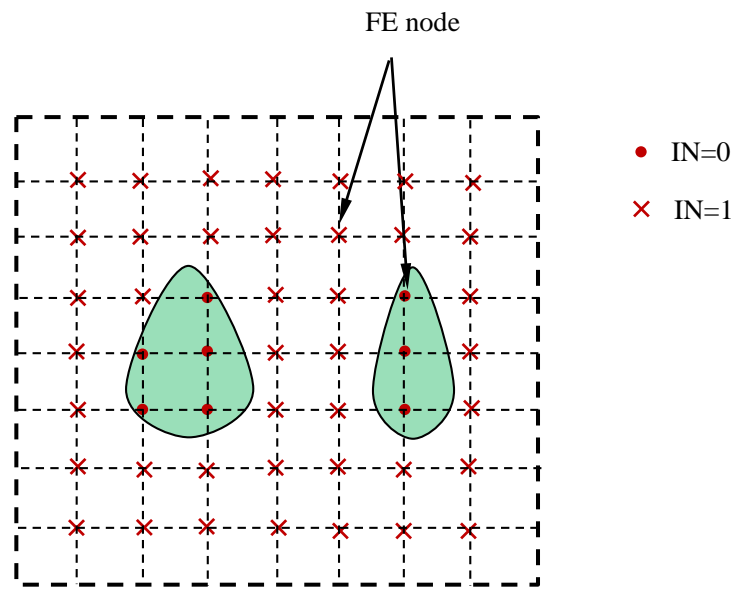

Fig. 13 A schematic illustration of the construction of $\chi^s$.

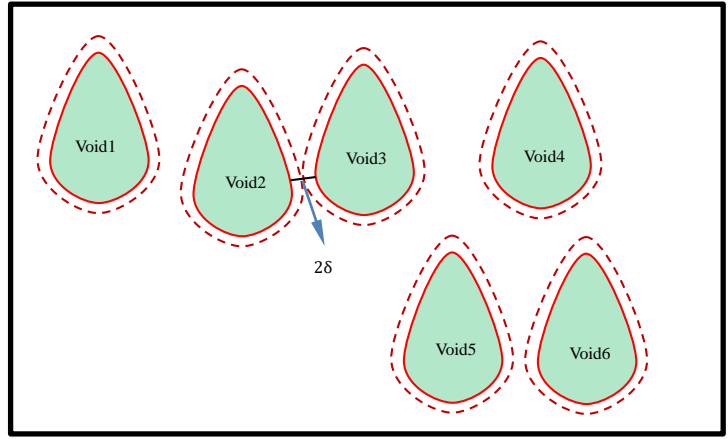

Fig. 14 A schematic illustration of the non-intersection constraint with a δ-expsnsion treatment.

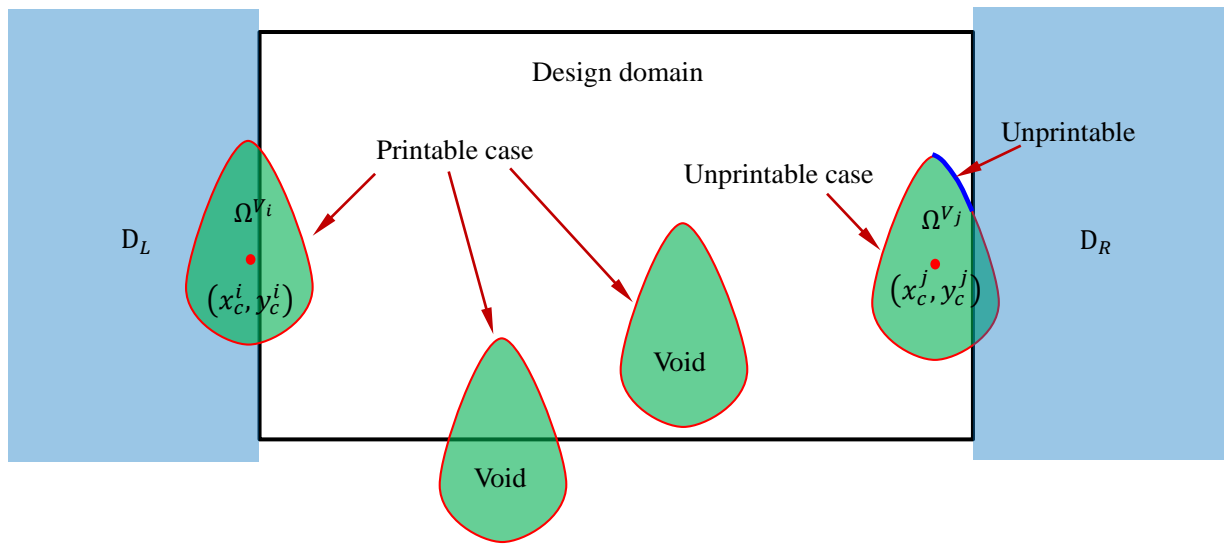

Fig. 15 A schematic illustration of the meanings of the symbols appeared in the Eq. (4.5).

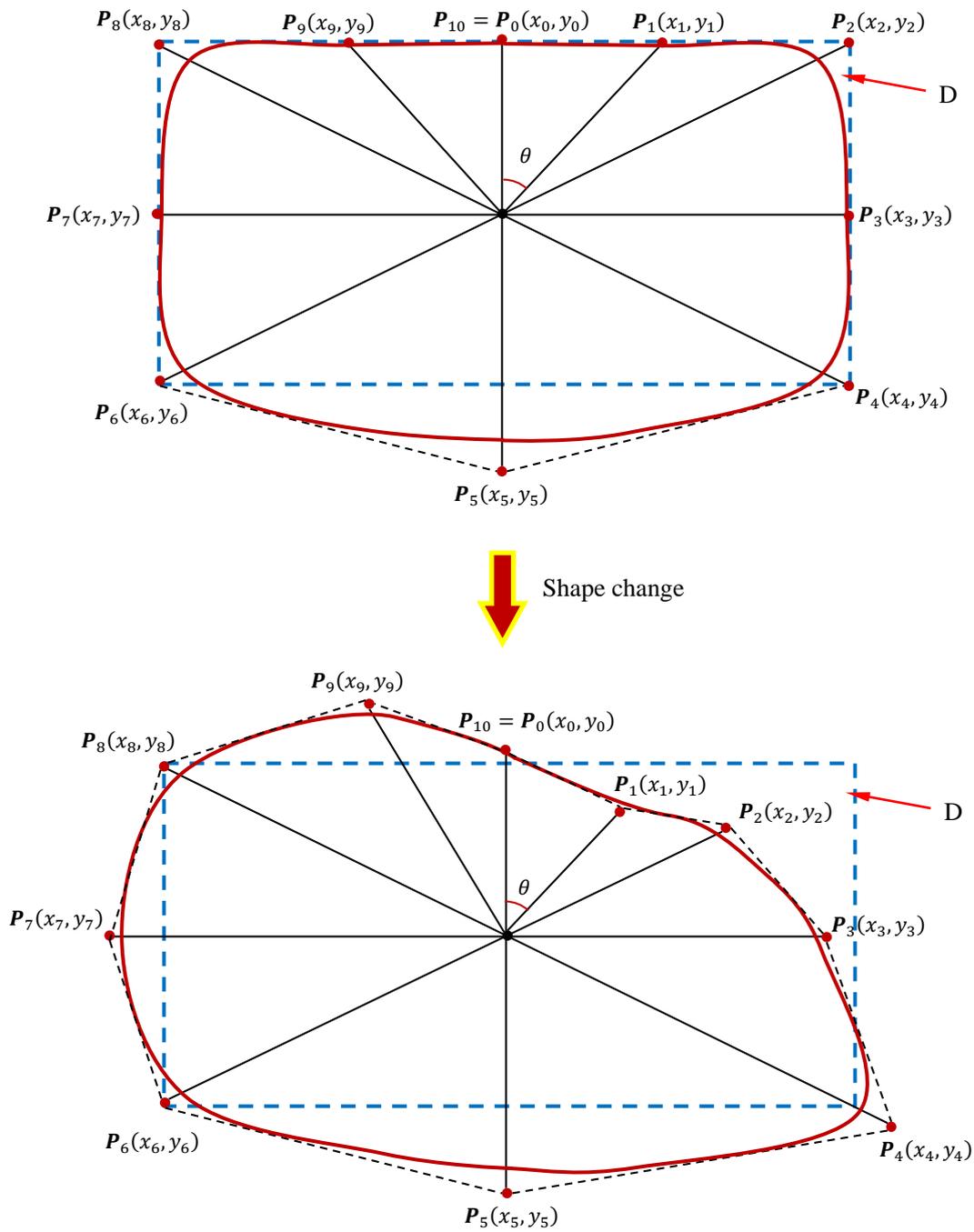

Fig. 16 A schematic illustration of the B-spline representation of the exterior boundary of a structure.

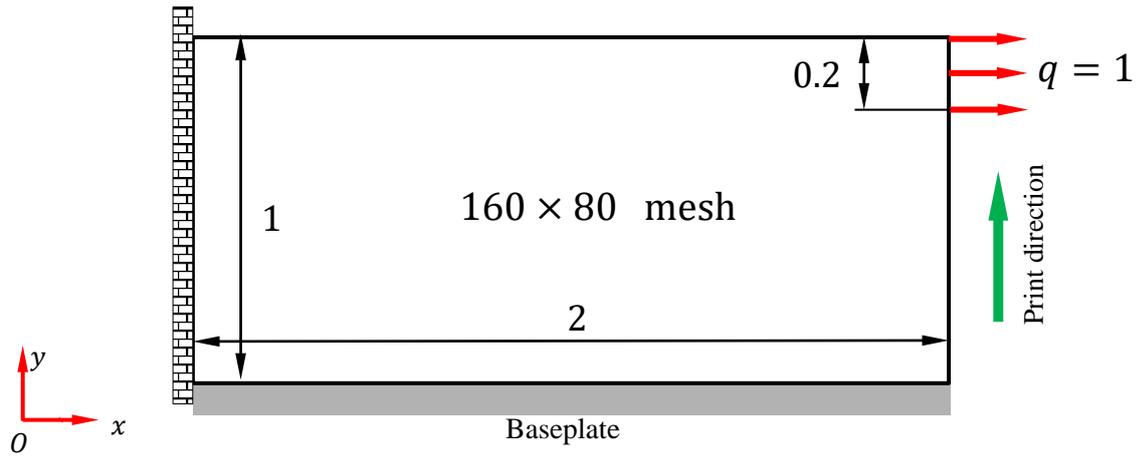

Fig. 17 The tensile beam example.

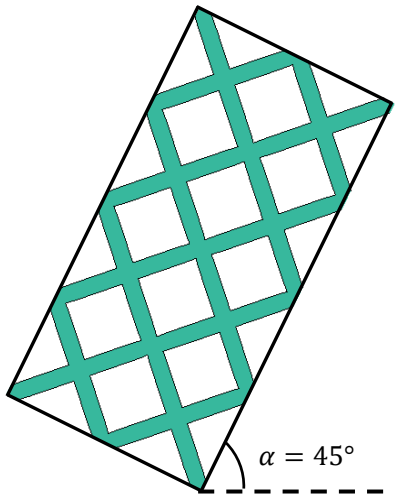

Fig. 18 The initial design of the tensile beam problem under the MMC-based formulation.

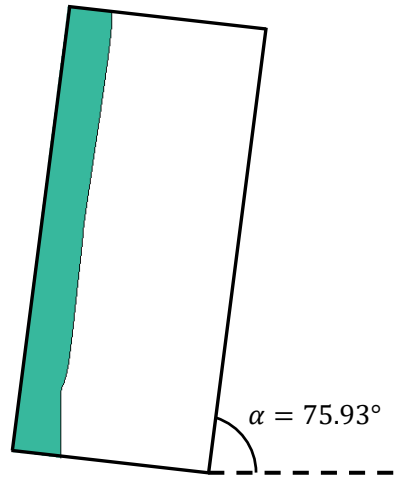

(a). The optimized structure (contour plot).

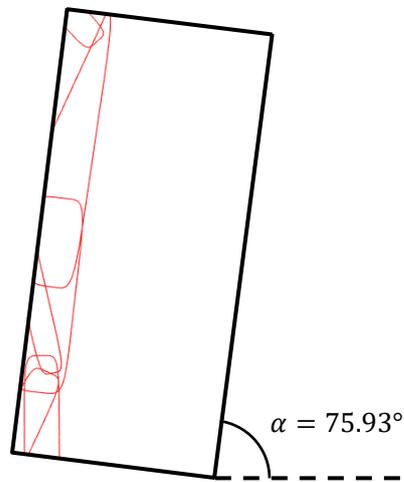

(b). The optimized structure (component plot).

Fig. 19 The optimized structure obtained with the MMC-based formulation.

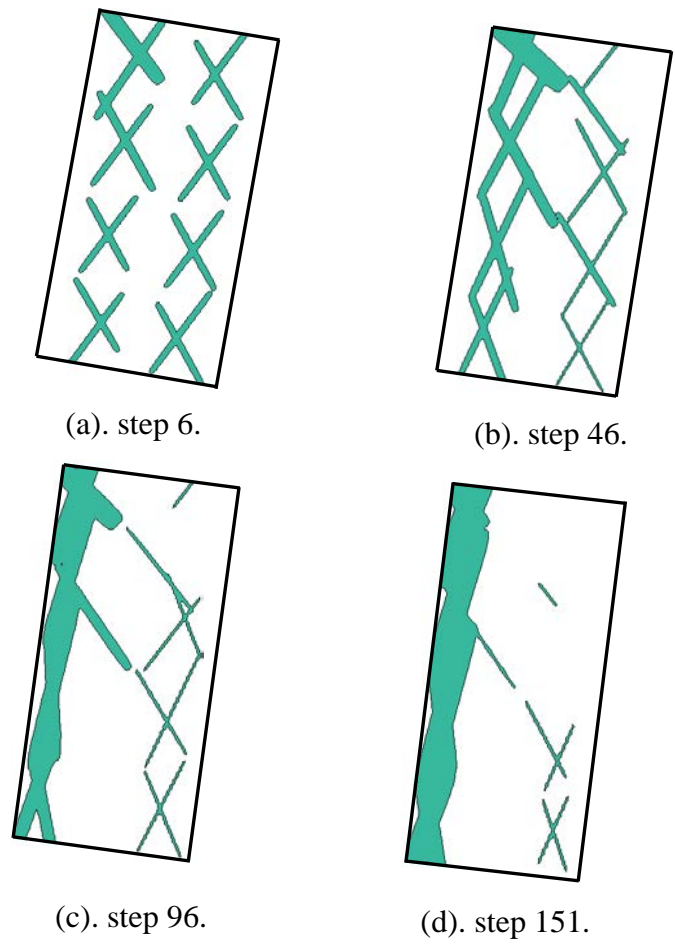

(a). step 6.   (b). step 46.

(c). step 96.   (d). step 151.

Fig. 20 Some intermediate steps of the optimization process.

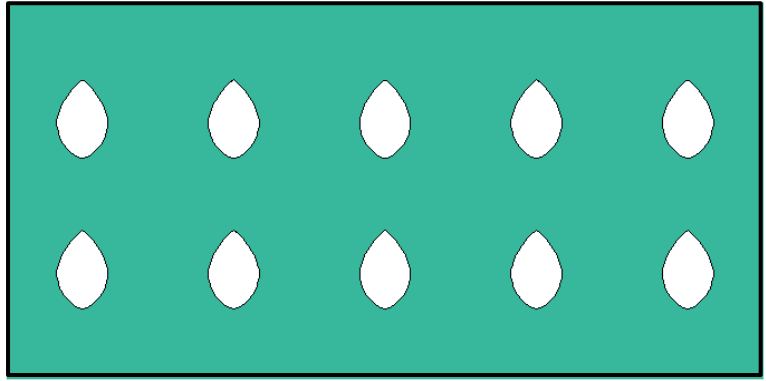

(a). A initial design with 10 voids.

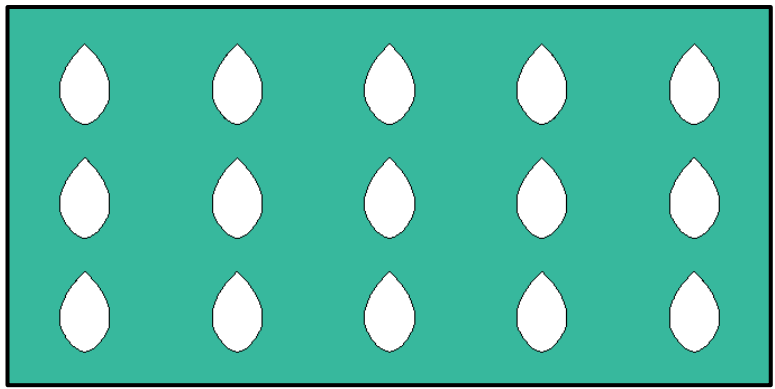

(b). A initial design with 15 voids.

Fig. 21 The initial designs of the tensile beam problem ($\boldsymbol{b}_p = (0,1)^\top$).

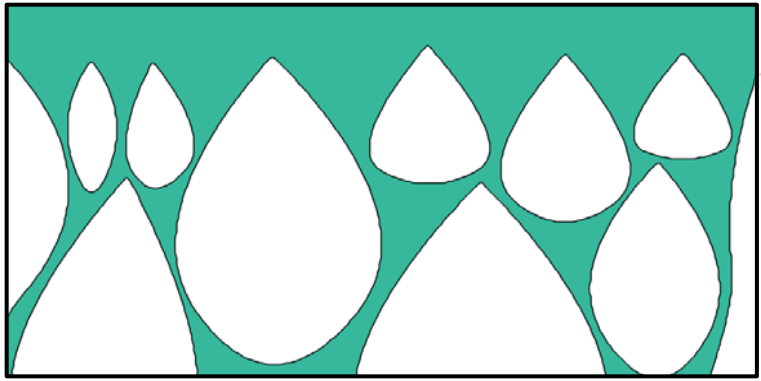

(a). The optimized structure obtained with initial design shown in Fig. 21a.

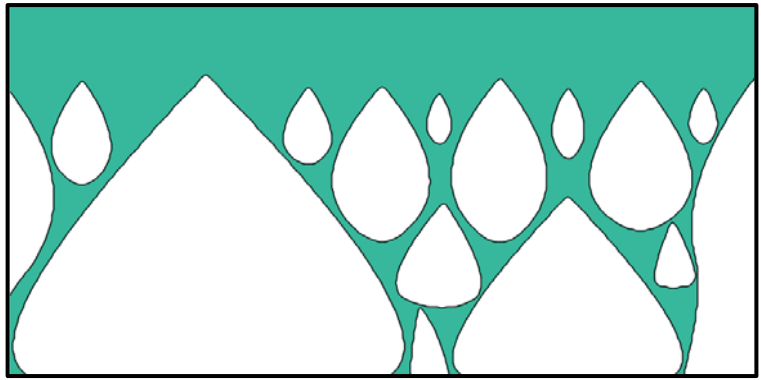

(b). The optimized structure obtained with initial design shown in Fig. 21b.

Fig. 22 The optimized structures of the tensile beam problem ($\boldsymbol{b}_p = (0,1)^\top$).

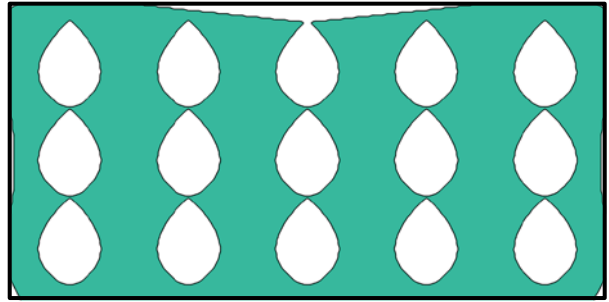

(a). step 2.

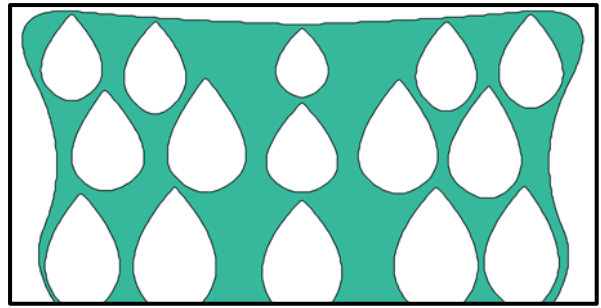

(b). step 20.

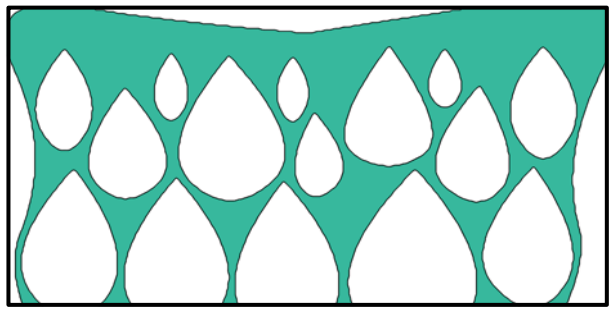

(c). step 95.

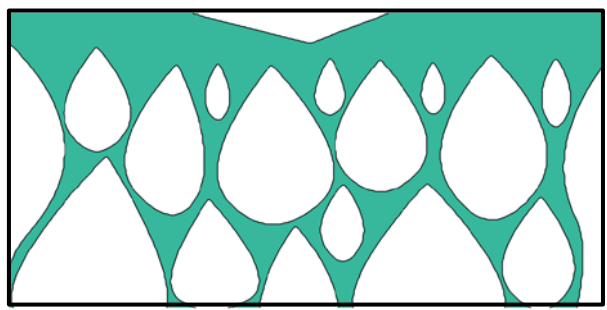

(d). step 465.

Fig. 23 Some intermediate steps of the optimization process with the initial design shown in Fig. 21b.

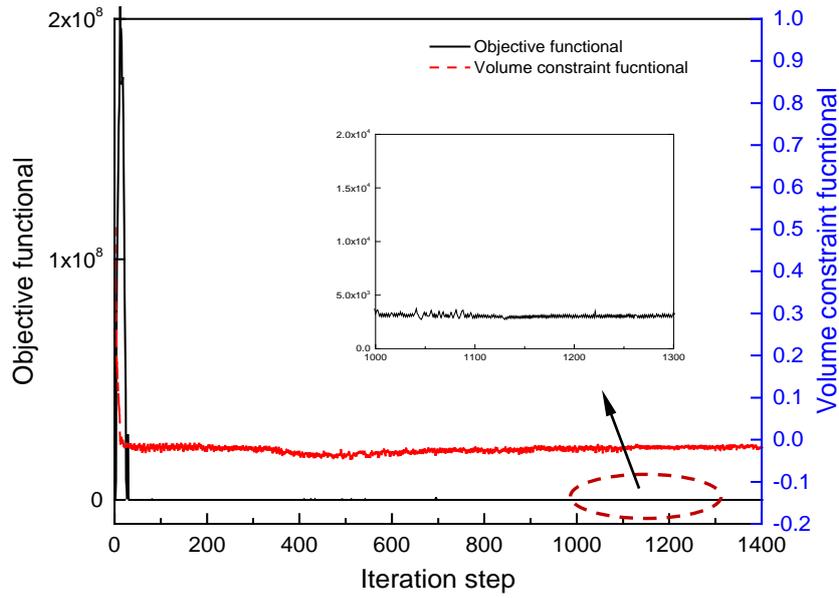

(a). Convergence history of the optimized structure with initial design shown in Fig. 21a.

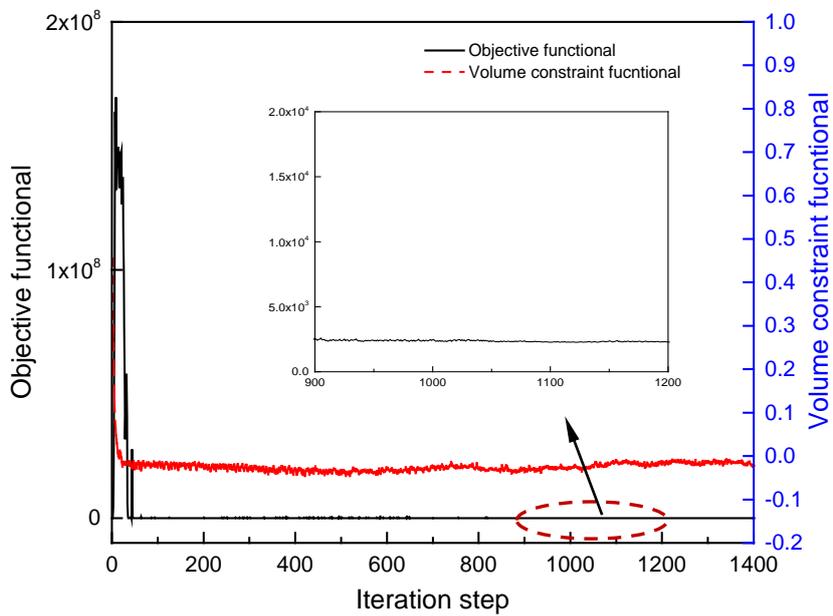

(b). Convergence history of the optimized structure with initial design shown in Fig. 21b.

Fig. 24 Convergence history of the tensile beam problem ($\boldsymbol{b}_p = (0,1)^\top$).

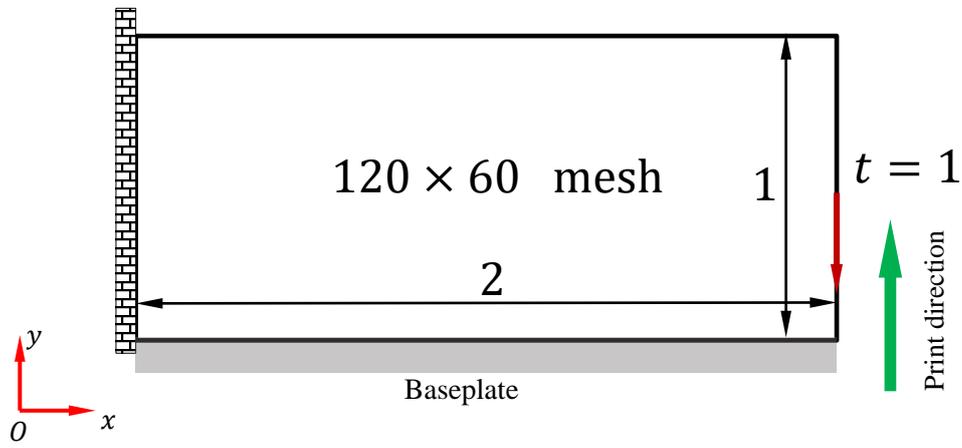

Fig. 25 The short beam example.

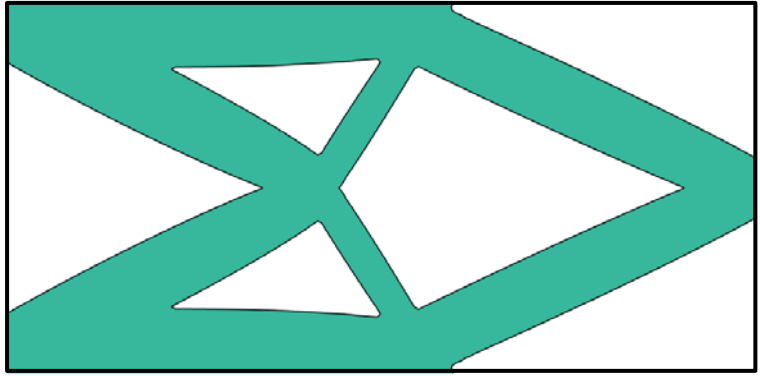

Fig. 26 The optimized structure without considering self-supporting requirement.

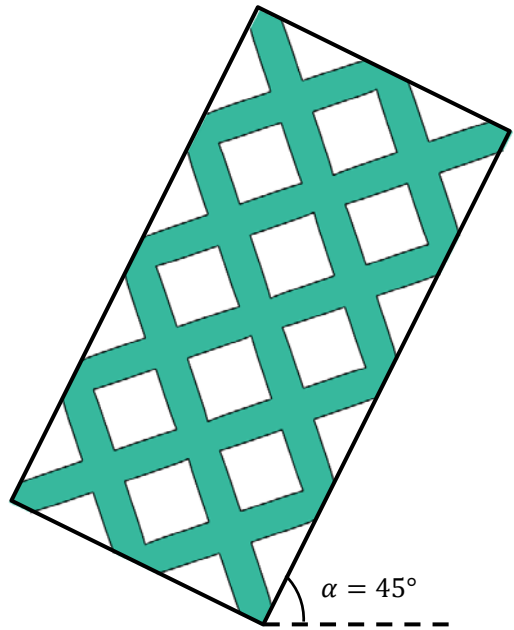

(a). A initial design with working plane $\alpha = 45°$.

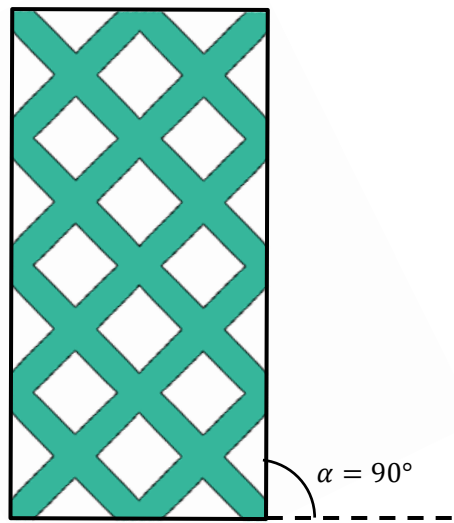

(b). A initial design with working plane $\alpha = 90°$.

Fig. 27 The initial designs of the short beam problem under the MMC formulation.

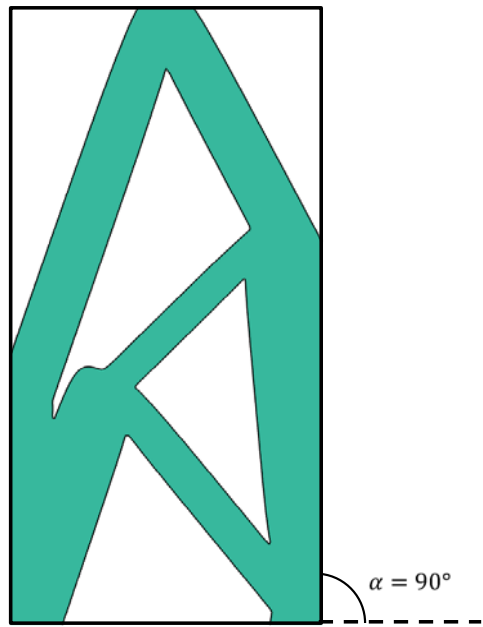

(a). The optimized structure (contour plot).

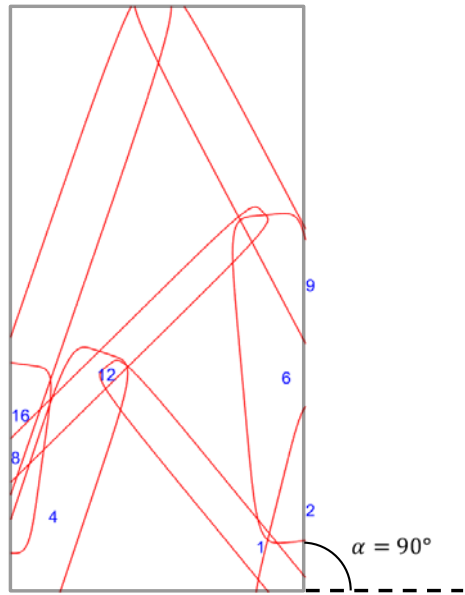

(b). The optimized structure (component plot).

Fig. 28 The optimized structure obtained with initial design shown in Fig. 27a

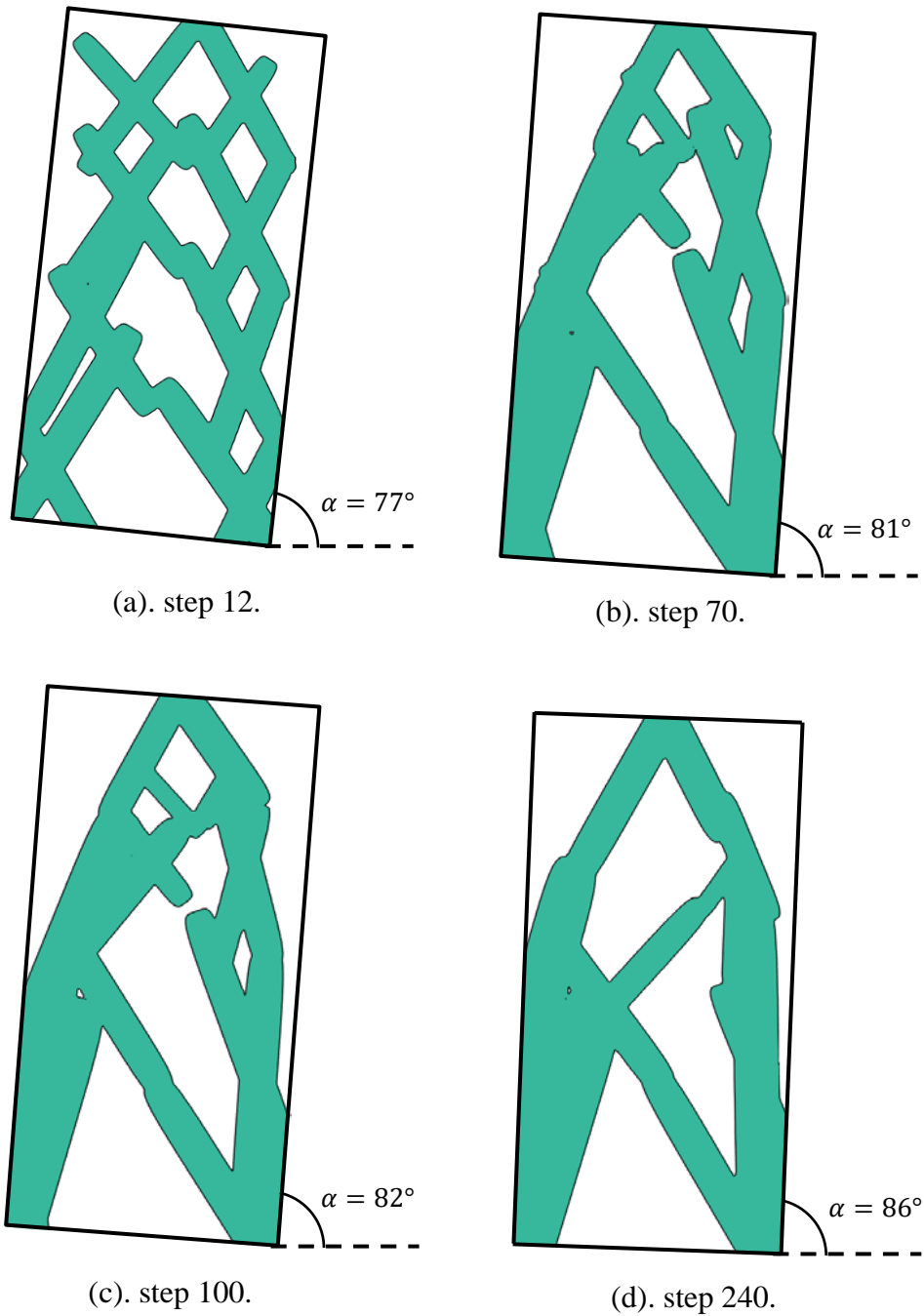

Fig. 29 Some intermediate steps of the optimization process with the initial design shown in Fig. 27a.

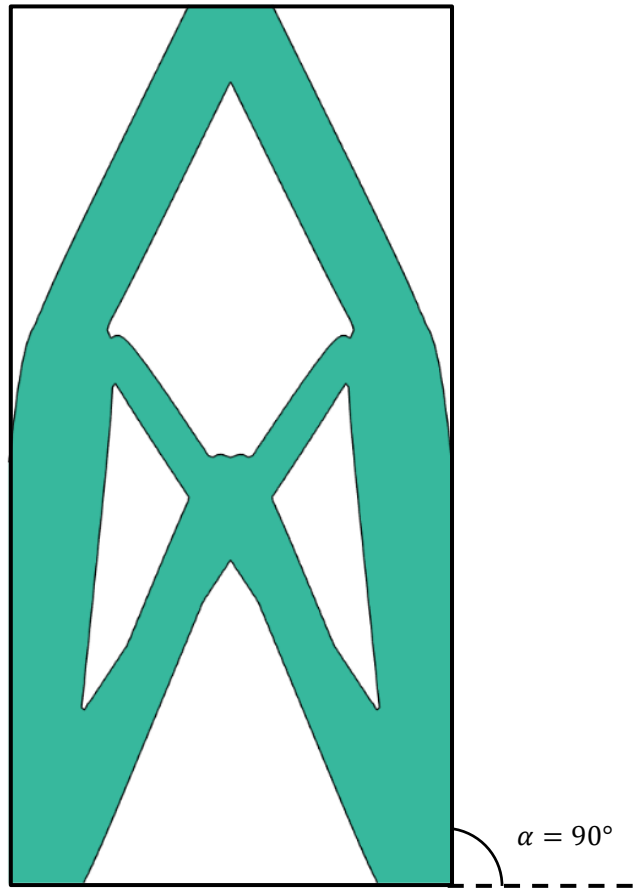

Fig. 30 The optimized structure obtained with the initial design shown in Fig. 27b under the MMC formulation.

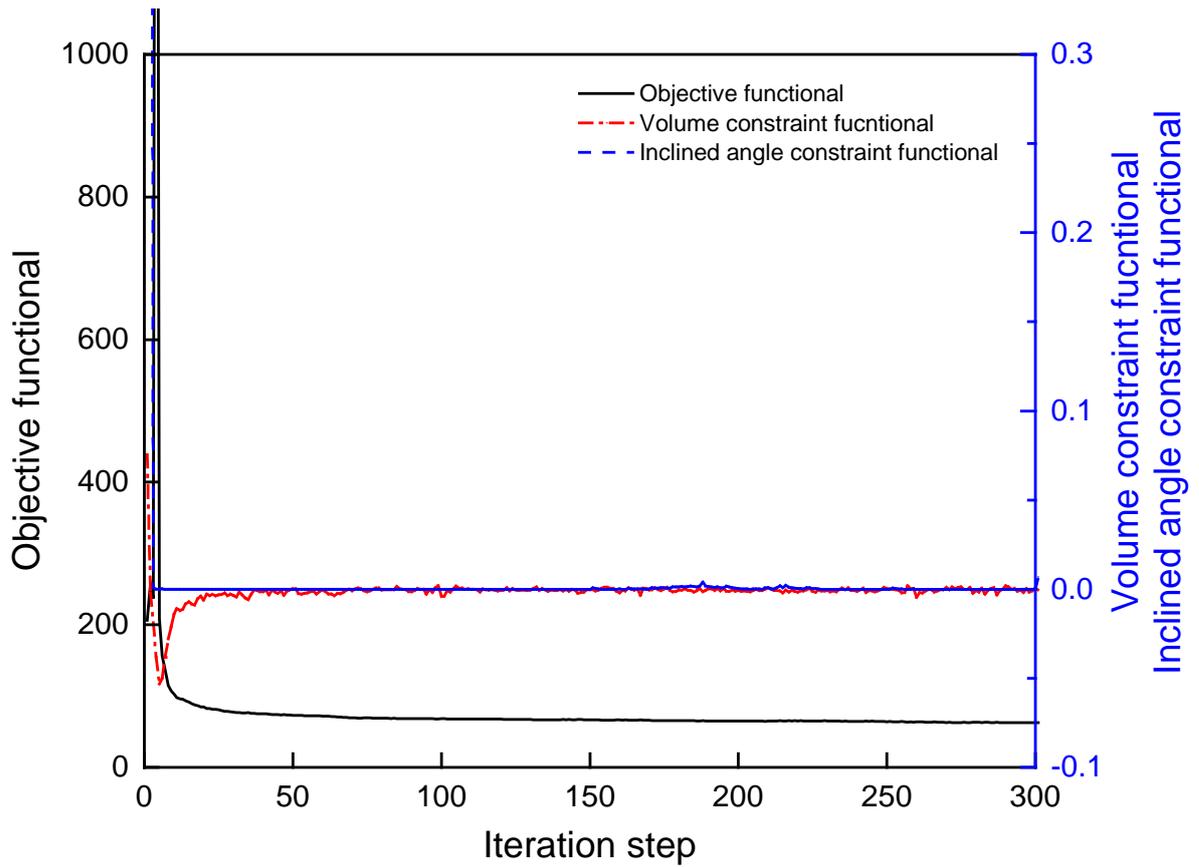

Fig. 31 Convergence history of the short beam example with the initial design shown in Fig. 27a under the MMC formulation.

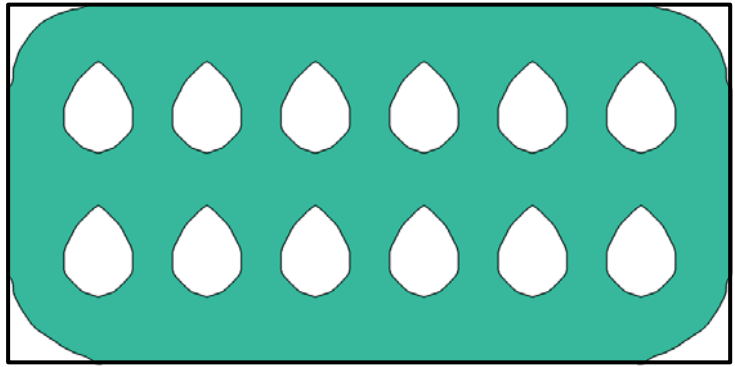

Fig. 32 The initial design of the short beam problem under the MMV formulation ($\boldsymbol{b}_p = (0,1)^\top$).

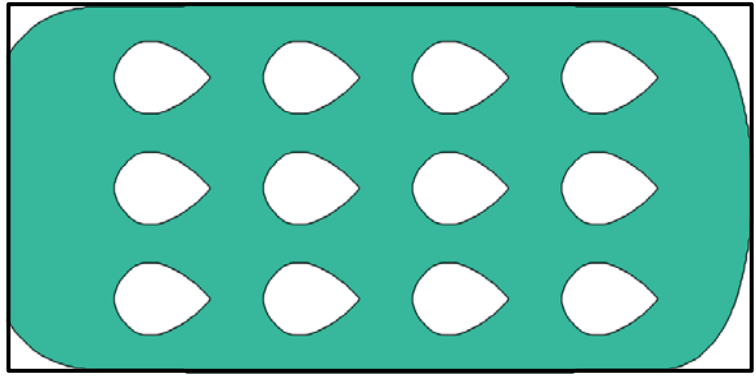

Fig. 33 The initial design of the short beam problem under the MMV formulation ($\boldsymbol{b}_p = (1,0)^\mathsf{T}$).

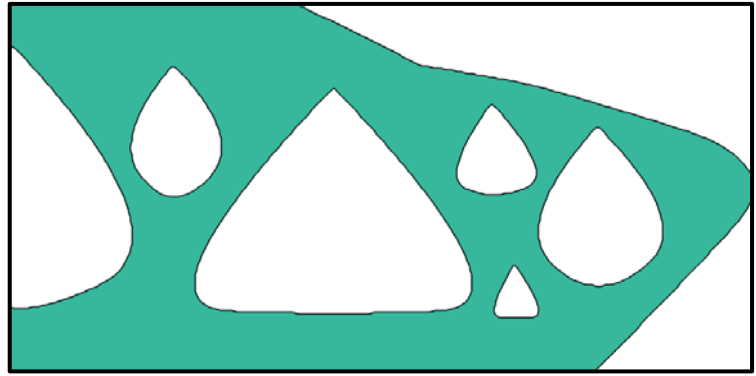

(a). The optimized structure (contour plot).

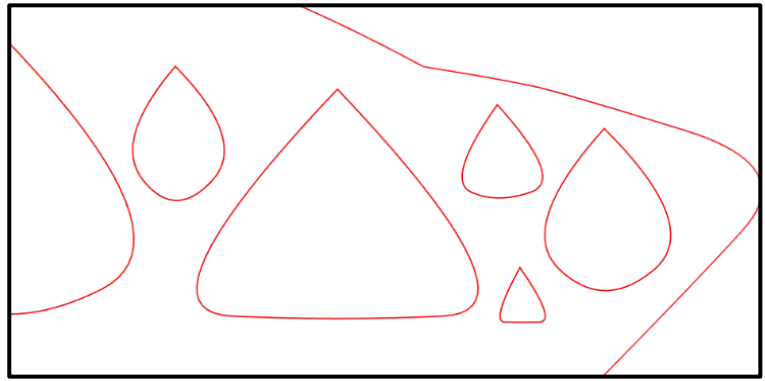

(b). The optimized structure (B-spline plot).

Fig. 34 The optimized structures obtained with the initial design shown in Fig. 32 under the MMV formulation ($\boldsymbol{b}_p = (0,1)^\top$).

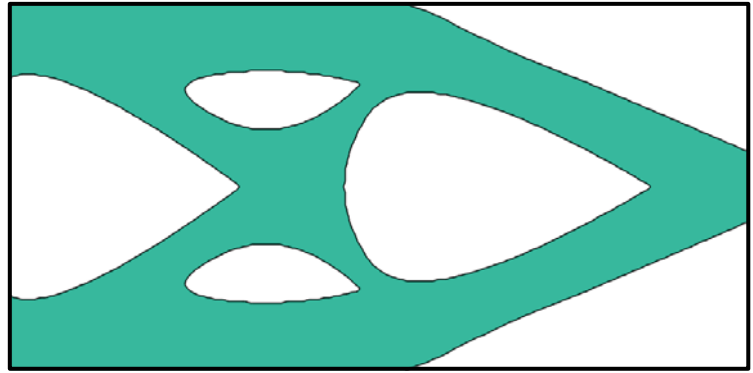

(a). The optimized structure (contour plot).

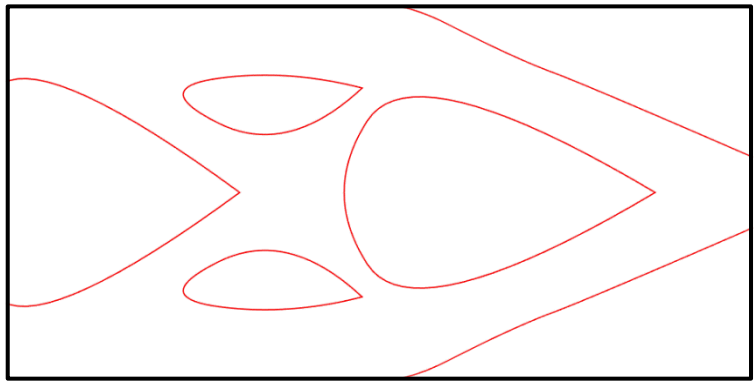

(b). The optimized structure (B-spline plot).

Fig. 35 The optimized structure obtained with the initial design shown in Fig. 33 under the MMV formulation ($\boldsymbol{b}_p = (1,0)^\top$).

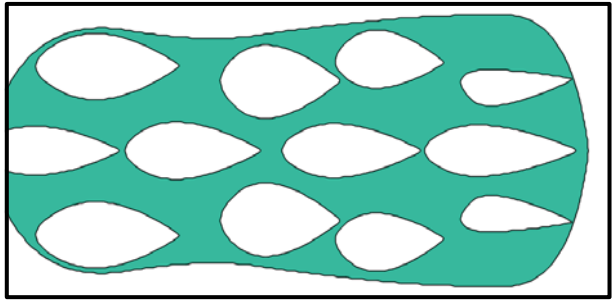

(a). step 7.

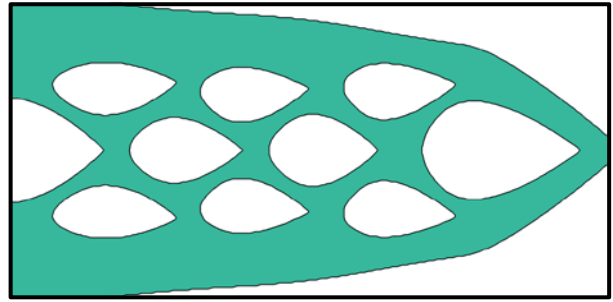

(b). step 103.

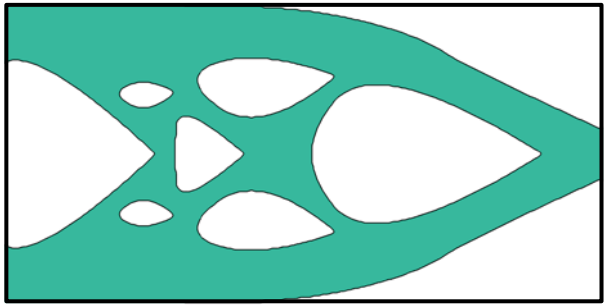

(c). step 1021.

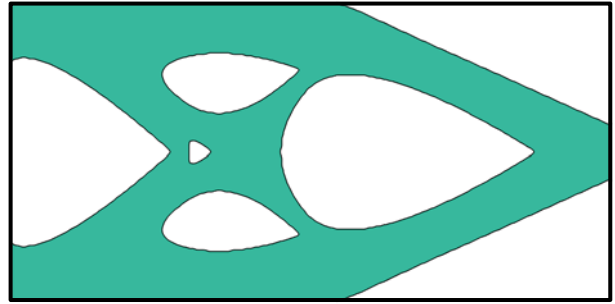

(d). step 1327.

Fig. 36 Some intermediate steps of the optimization process with the initial design shown in Fig. 33 ($\boldsymbol{b}_p = (1,0)^\top$).

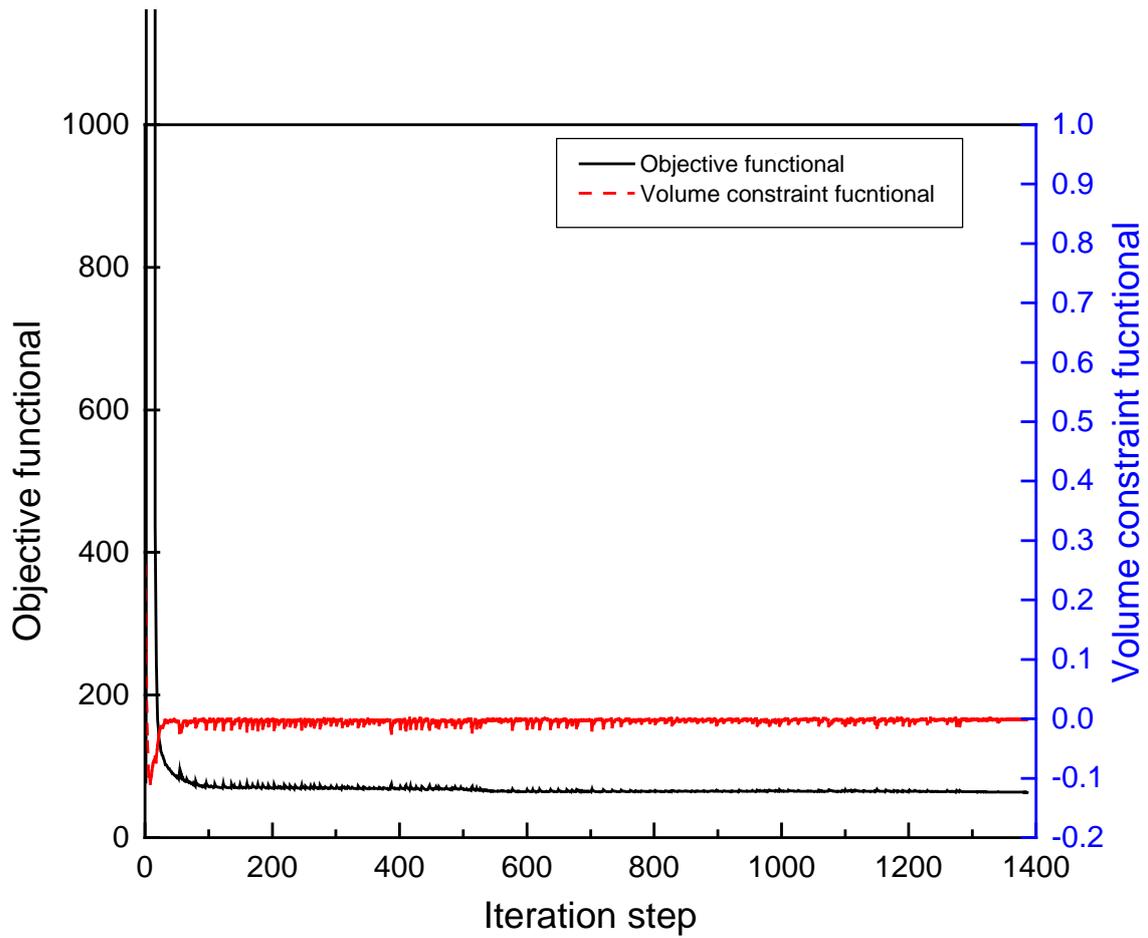

Fig. 37 Convergence history of the short beam example with initial design shown in Fig. 33 under the MMV formulation..

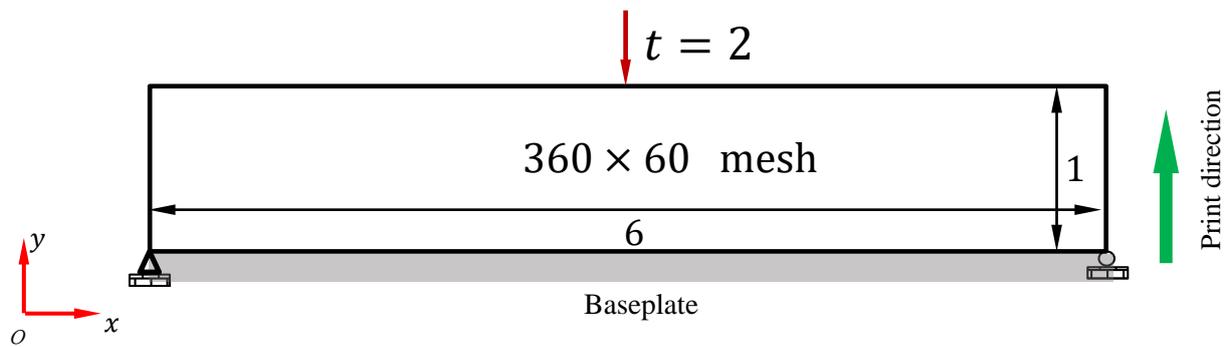

Fig. 38 The MBB example.

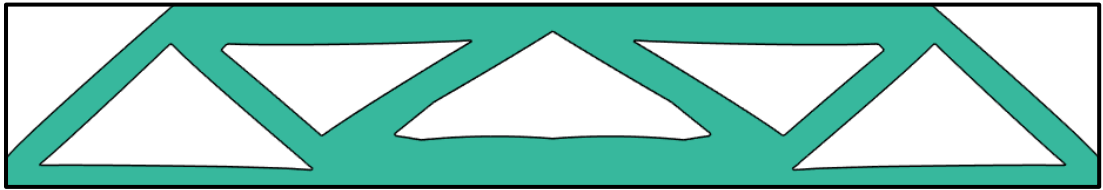

Fig. 39 The optimized structure without considering self-supporting requirement.

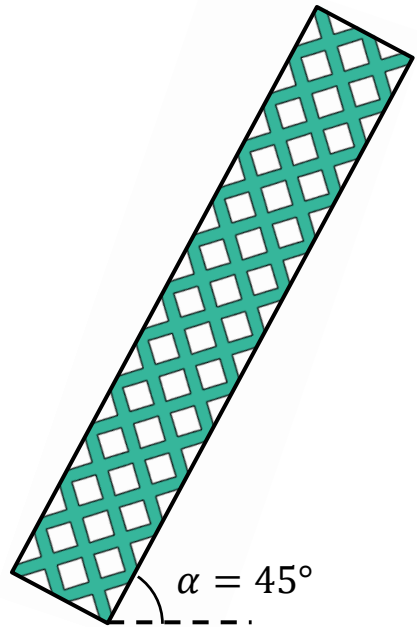

Fig. 40 The initial design of the MBB problem under the MMC-based formulation.

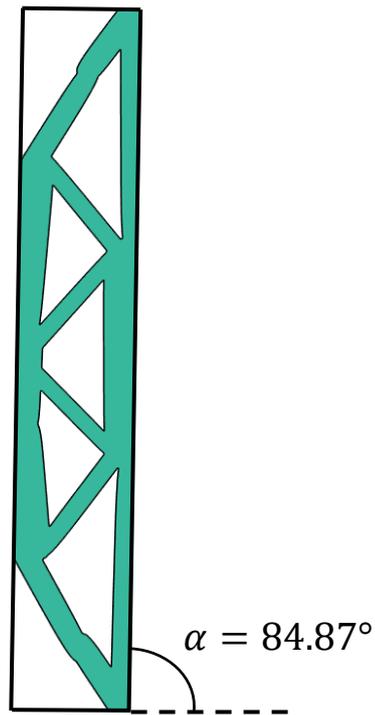

(a). The optimized structure (contour plot).

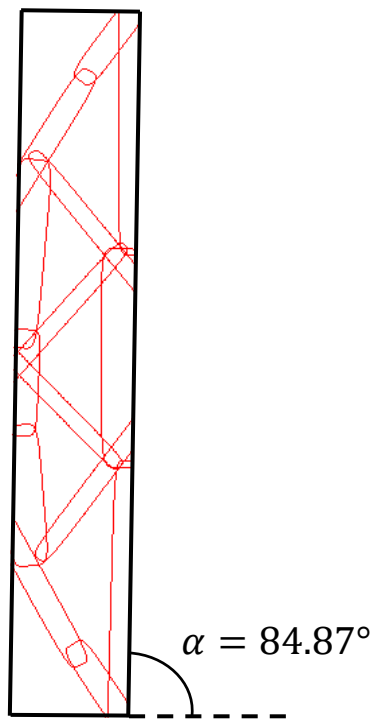

(b). The optimized structure (component plot).

Fig. 41 The optimized structure obtained with the MMC-based approach.

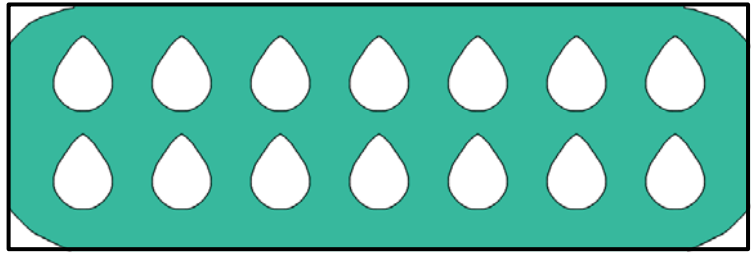

Fig. 42 The initial design of the MBB problem under the MMV-based formulation ($\boldsymbol{b}_p = (0,1)^\top$).

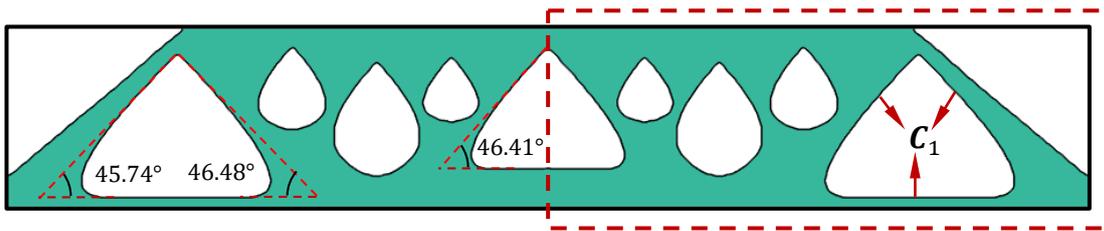

(a). Optimized solution considering self-supporting constraint.

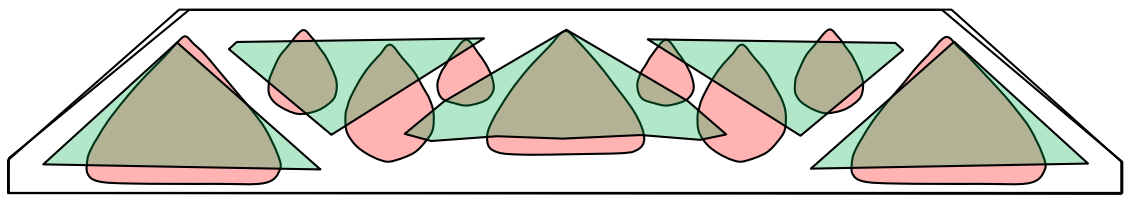

(b). A comparison with the unconstrained optimized solution.

Fig. 43 The optimized structure obtained with the MMV-based approach ($\boldsymbol{b}_p = (0,1)^\top$).

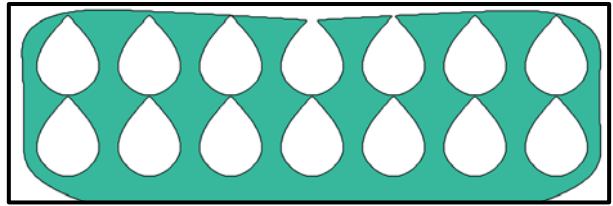

(a). step 2.

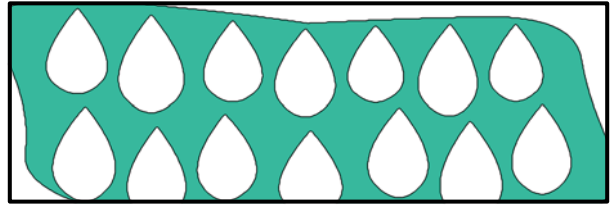

(b). step 43.

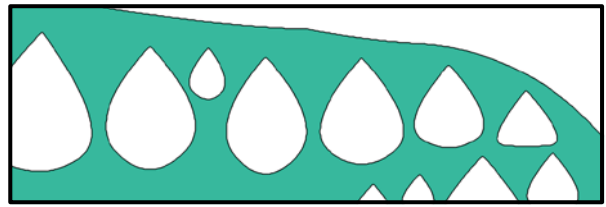

(c). step 337.

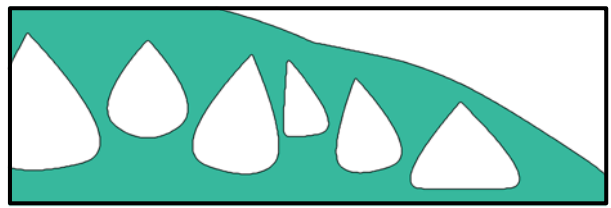

(d). step 1072.

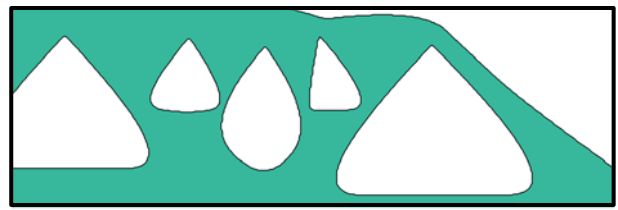

(e). step 1555.

Fig. 44 Some intermediate steps of the optimization process.

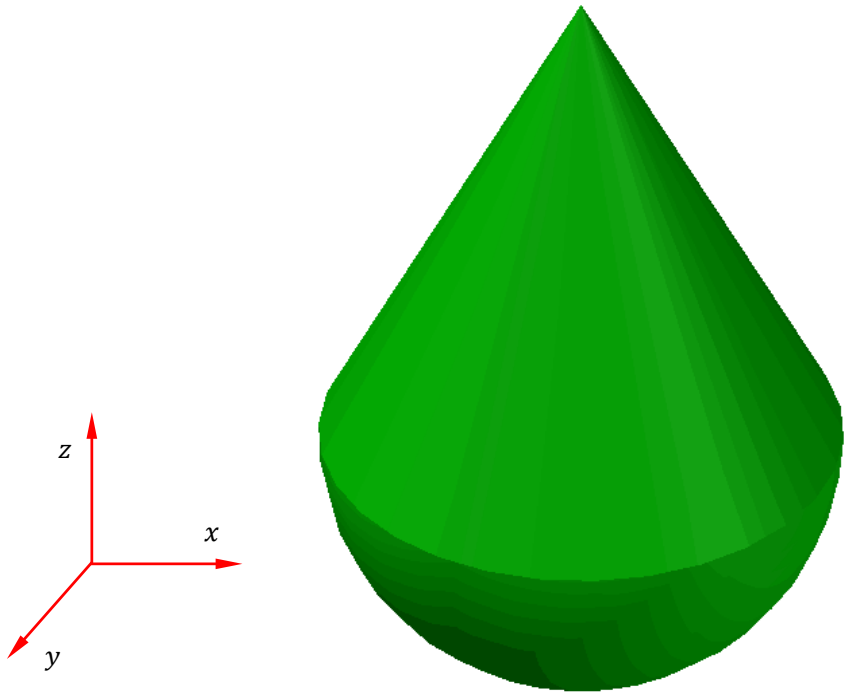

Fig. 45 A schematic illustration of a 3D printable feature.

Tab. 1 Optimal solution of the short beam example.

| The number of active component | Inclined angle (local coordinate system) | Inclined angle (global coordinate system) |
|---|---|---|
| 1 | 39.18° | 129.18° |
| 2 | -14.02° | 75.98° |
| 4 | -18.49° | 71.51° |
| 6 | 5.13° | 95.13° |
| 8 | -7.97° | 82.03° |
| 9 | 27.74° | 117.74° |
| 12 | -44.23° | 45.77° |
| 16 | -19.04° | 70.96° |